\documentclass[a4paper,11pt]{article}
\usepackage{jheppub}

\usepackage{subfigure}
\usepackage{float}
\usepackage{color}
\usepackage[usenames,dvipsnames,svgnames,table]{xcolor}
\usepackage{amssymb,amsmath,amsfonts}
\usepackage{braket} %please add this package we need it.
\usepackage{upgreek}
\usepackage{bbm,bm}

\DeclareMathSizes{10}{10}{6}{6}
\title{Chiral Magnetic Effect and Three-point Function from AdS/CFT Correspondence}
\author[a,b]{Lei Yin \footnote{ \href{mailto: lei@scnu.edu.cn}{lei@scnu.edu.cn}}}
\author[c]{Defu Hou \footnote{Co-corresponding author: \href{mailto: houdf@mail.ccnu.edu.cn}{houdf@mail.ccnu.edu.cn}}}
\author[c,d]{Hai-cang Ren \footnote{Co-corresponding author: \href{mailto: renhc@mail.ccnu.edu.cn}{renhc@mail.ccnu.edu.cn}}}

\affiliation[a]{Guangdong Provincial Key Laboratory of Nuclear Science,
Institute of Quantum Matter, South China Normal University, Guangzhou 510006, China}

\affiliation[b]{Guangdong-Hong Kong Joint Laboratory of Quantum Matter, Southern Nuclear Science Computing Center, South China Normal University, Guangzhou 510006, China}

\affiliation[c]{Institute of Particle Physics and Key Laboratory of Quark and Lepton Physics (MOE), \\ Central China Normal University, Wuhan 430079, China}

\affiliation[d]{Physics Department, The Rockefeller University, 1230 York Avenue, New York,10021-6399, USA}

\abstract{
  The chiral magnetic effect with a fluctuating chiral imbalance is more realistic in the evolution of quark-gluon plasma, which reflects the random gluonic topological transition. Incorporating this dynamics, we calculate the chiral magnetic current in response to  space-time dependent axial gauge potential and magnetic field in  AdS/CFT correspondence. In contrast to conventional treatment of constant axial chemical potential, the response function here is the AVV three-point function of the $\mathcal{N}=4$ super Yang-Mills at strong coupling. Through an iterative solution of the nonlinear equations of motion in Schwarzschild-AdS$_5$ background, we are able to  express the AVV function in terms of two Heun functions and prove its UV/IR finiteness, as expected for $\mathcal{N}=4$ super Yang-Mills theory.  We found that the dependence of the chiral magnetic current on a non-constant chiral imbalance is non-local, different from hydrodynamic approximation, and demonstrates the subtlety of the infrared limit discovered in field theoretic approach. We expect our results enrich the understanding of the phenomenology of the chiral magnetic effect in the context of relativistic heavy ion collisions.
}


\begin{document}

\maketitle

\tableofcontents
\flushbottom

%%%%%%%%%%%%%%%%%%%%%%%%%%%%%%%%%%%
\section{Introduction and Summary}
\label{sec:introduction}

    Chiral matter subject to an external magnetic field and/or under rotation exhibits many interesting transport properties driven by the axial anomaly. Among them are the chiral magnetic effect and chiral vortical effect \cite{CME1, CME2, CME3, CVE}.  Searching the evidences of these novel anomalous transport phenomena has grown into an active research area for the past decade with the scope extending from the quark-gluon plasma (QGP) created in relativistic heavy ion collision \cite{CMEinSTAR, CMEinALICE, CMEinCMS} to the Weyl semi-metals \cite{Son_Spivak,CMEinSemimetals_1,CMEinSemimetals_2,Katsuhisa_Taguchi} and involved both theoretician and experimentalist. This theoretical work focuses on the chiral magnetic effect in QGP. 

The chiral imbalance in QGP is triggered by the topological excitation of QCD and the external magnetic field is produced via the off-central collision of heavy ions. The resulting chiral anomaly is reflected in the anomalous Ward identity of the axial-vector current $J_{\rm A}^\mu$ in the presence of vector and axial vector field strengths 
$( F^{\mathrm{V}})_{\mu\nu}$ and $(F^{\mathrm{A}})_{\mu\nu}$
\begin{equation}
  \partial_\mu J_{\mathrm{A}}^\mu(x) = \mathcal{C} \epsilon^{\mu \nu \rho \sigma} \bigg[ \, (F^{\mathrm{V}})_{\mu\nu} ( F^{\mathrm{V}})_{\rho \sigma} + \frac{1}{3} ( F^{\mathrm{A}})_{\mu\nu} ( F^{\mathrm{A}})_{\rho \sigma} \bigg] \;,
  \label{anomaly}
\end{equation}
and the chiral magnetic current for constant axial chemical potential $\mu_A$ and magnetic field $\bm{B}$ takes the simple form
\begin{equation}
  \bm{J} = 8 \mathcal{C}\; \mu_A\bm{B} \;,
  \label{CME}
\end{equation}
with the non-renormalization anomaly coefficient $\mathcal{C}$. Here, the constant axial chemical potential serves as the Lagrange multiplier of a grand canonical ensemble of macroscopic chirality. The chiral magnetic effect is thereby a direct probe of the topological structure of QCD, more important than other anomalous transport phenomena in this sense.

Theoretically, the chiral magnetic effect has been investigated in different approaches, including the Green function formalism, \cite{CME3, Hou}, kinetic theories \cite{JHGao}, and holography \cite{Yee2009, Gynther2011, Rebhan, Bu2016a}. Most of these works focus on the situation with a (nearly) constant $\mu_A$. Hydrodynamic simulations have also been developed for RHIC, based on the assumption that a net axial charge density is generated in the initial stage of collisions and its characteristic time of variation is much longer than the relaxation time to thermal equilibrium \cite{YJiang,Shi,Guo}.  

The chiral magnetic response in the non-equilibrium case, in particular for a spacetime-dependent chiral imbalance and magnetic field, turns out to be both subtle and more realistic for heavy-ion-collisions. The initial axial charge is generally expected to be inhomogeneous across the fireball and furthermore necessarily evolves in time due to random gluonic topological transitions during the fireball evolution. The magnetic field generated during the collision is also transient. The spatial variation length scale and the time evolution scale are not necessarily very large as compared with the thermal scale of the medium. Exploring the dynamics of the chiral magnetic effect under inhomogeneous and non-static magnetic field and chiral imbalance is the main target of the present work. To simulate the strong coupling feature of the QGP created in RHIC, the AdS/CFT correspondence is employed with the $\mathcal{N}=4$ super-Yang-Mills at large number of colors and large `t Hooft coupling and its global $U(1)$ vector current as the proxies of QCD in deconfinement phase and electric current. 

In the presence of an axial chemical potential, an external vector potential $\mathsf{V}_\mu(q_1)$ and an axial vector potential $\mathsf{A}_\mu(q_2)$, the response current in 4-momentum representation can be expanded according to the powers of the external fields, i. e.
\begin{align}
  J^\mu(q) = \Pi^{\mu\nu}(q)\mathsf{V}_\nu(q) + J_{\rm CME}^\mu(q) + ... 
\end{align}
where the first term, linear in electromagnetic field, represents the ordinary polarization current and the second term, bilinear in electromagnetic field and axial vector potential,  gives rise to the chiral magnetic effect to be discussed in this work. The ellipsis represents terms in higher powers of the external fields. The chiral magnetic current $J_{\rm CME}^\mu(q)$ can be divided into two terms, i.e.
\begin{equation}
  J_{\rm CME}^\mu(q) = \mathsf{J}^\mu(q) + J_{\rm AVV}^\mu(q) \;. \label{eq:52}
\end{equation}
The first term consists of only spatial component, $\mathsf{J}^\mu(q) = \{0, \bm{\mathsf{J}}(q) \}$
\begin{equation}
  \bm{\mathsf{J}}(q) = \mu_A \, K(q)\,\bm{\mathsf{B}}(q) \label{eq:21}
\end{equation} 
that generalizes Eqn.(\ref{CME}) to arbitrary spacetime-dependent magnetic field at a constant axial chemical potential with the explicit form of the kernel $K(q)$ given by (\ref{kernelK}) in Section~\ref{sec:smallq}. The second term
\begin{eqnarray}
  J_{\rm AVV}^\mu(q) = \int \frac{\mathrm{d}^4q_1}{(2\pi)^4} \frac{\mathrm{d}^4q_2}{(2\pi)^4} (2\pi)^4 \delta^4(q_1+q_2-q)
  \Lambda^{\mu\nu\rho}(q_1,q_2)\mathsf{A}_\rho(q_2)\mathsf{V}_\nu(q_1)
  \label{weak}
\end{eqnarray}
brings in the spacetime dependence of the chiral imbalance with the integration kernel $\Lambda^{\mu\nu\rho}(q_1,q_2)$ related to the AVV three point function with $q_1=(\omega_1,\bm{q}_1)$ and $q_2=(\omega_2,\bm{q}_2)$ the 4-momenta carried by a vector gauge field $\mathsf{V}_\nu$ and an axial gauge field $\mathsf{A}_\nu$, respectively. In terms of the standard notation of the AVV three point function
\begin{equation} 
  \Delta^{\mu\nu\rho}(k_1,k_2) = \langle  J_V^\mu(k_1)J_V^\nu(k_2)J_A^\rho(-k_1-k_2)\rangle \;,  \label{eq:32}
\end{equation}
with $J_V$ and $J_A$ the vector and axial-vector currents operators
\begin{equation}
  \Lambda^{\mu\nu\rho}(q_1,q_2) = \Delta^{\mu\nu\rho}(-q_1-q_2,q_1)  \;. \label{eq:34}
\end{equation}
In what follows, we shall mainly focus on the AVV contribution to the chiral magnetic current. To our knowledge, the kernel $\Lambda^{\mu\nu\rho}(q_1,q_2) $ with both $q_1$ and $q_2$ nonzero have not been examined in strong coupling.

\begin{figure}[!htb]
  \centering{\includegraphics[width=2.3in]{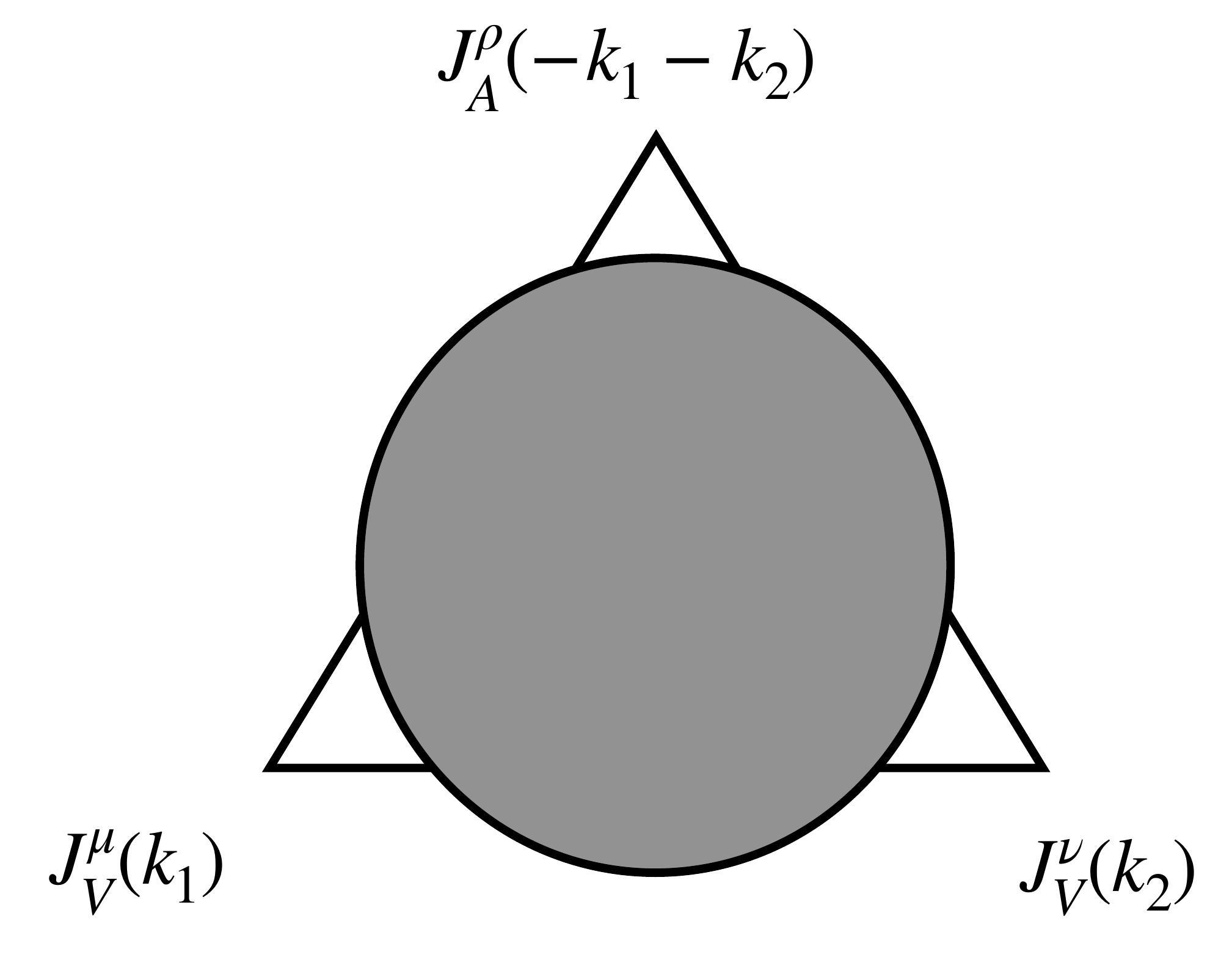}}
  \caption{ The triangle diagram of AVV three-point function, where the shaded center area implies strong-coupling, contrasting with one-loop weakly-coupled counterpart.}
  \label{fig:triangle}
\end{figure} 

Because of the anomaly, $\mu_A$ cannot be identified with $\mathsf{A}_0$ \cite{Rubakov2010,Gynther2011}. Physically, $\mu_A$ is conjugate to a conserved global axial-charge and is thereby a constant in an equilibrium. The spacetime variation of the chiral imbalance is attributed to $\mathsf{A}_0$\footnote{In the absence of the spatial component of the axial vector potential, the gradient of $\mathsf{A}_0(q_2)$ gives rise to an axial-electric field, i.e. $\bm{\mathsf{E}}_A=i\bm{q}_2\mathsf{A}_0(q_2)$. The AVV three-point function contains the linear response to all orders of derivatives of $\bm{\mathsf{E}}_A$. Physically, however, the axial vector potential is not a dynamic degree of freedom. $\mathsf{A}_0$ is introduced merely to proxy arbitrary spacetime-dependent fluctuation of the axial charges because of the topological transitions of QCD that accompanies the evolution of the QGP fireballs.}. To distinguish their roles mathematically, we impose the condition
\begin{align}
  \mathsf{A}_0(q) \bigg|_{\bm{q}=0} \equiv \int \mathrm{d}^3\bm{x}\mathsf{A}_0(x) = 0 \;. \label{eq:55}
\end{align}
From holographic perspective, both $\mu_A$ and $\mathsf{A}_0(q)$ pertain to the temporal component of the axial vector potential {\it in the bulk} with its value at the horizon equal to $\mu_A$ and its value on the boundary equal to $\mathsf{A}_0(q)$. Unlike the hydrodynamic approach where the space-time variation of the chiral imbalance is treated as higher orders and thereby $|\mu_A| \gg |\mathsf{A}_0(x)|$, we consider $|\mu_A| \sim |\mathsf{A}_0(x)|$ throughout this work in order to investigate the non-equilibrium of chiral imbalance. Anticipating the stochastic nature of the topological transitions, $\mathsf{A}_0(q)$ may not be continuous in $q$, in particular, $\mathsf{A}_0(q)$ may not be small for a small but nonzero $\bm{q}$.

To calculated the chiral magnetic current in the super-Yang-Mills via AdS/CFT correspondence, we start with Einstein-Maxwell-Chern-Simons action in the $AdS_5$ bulk and solve the classical equations of motion in the background of a Schwarzschild black hole up to the first order of non-linearity in external gauge potentials. With equal order of magnitude of $\mu_A$ and $\mathsf{A}_\mu$, the metric fluctuation does not contribute to the terms displayed in the expansion (\ref{weak}), which implies that the spacetime dependence of temperature as well as $\mu_A$ can be ignored in evaluating the current to the displayed order. The weak external field approximation employed here corresponds to the physical condition $T \gg \mu_A$ and $T \gg \sqrt{e|\bm{\mathsf{B}}|}$ with $e = \mathcal{O}(\sqrt{\mathcal{C}})$ the electromagnetic coupling. The latter condition holds for the temperature inside the QGP phase but only marginally at the verge of the deconfinement transition.  Under this simplifications, we are able to develop an analytic formulation of $\Lambda^{\mu j0}(q_1,q_2)$ for arbitrary $q_1$ and $q_2$ in terms of two Heun functions, one of which reduces to a hypergeometric function for a homogeneous magnetic field. 

For low momenta $q$  ($|\bm{q}| \ll T$ and $|\omega| \ll T$), the kernel $K(q)$ in \eqref{eq:21} approaches to a constant
\begin{align}
  K (q) = 8\ \mathcal{C}
  \label{kernel_cme}
\end{align}
recovering the prototype CME formula (\ref{CME}) and the limit $q\to 0$ shows no ambiguity as demonstrated in \cite{Yee2009} some years ago. A derivative expansion with respect to the magnetic field can be developed as in the hydrodynamic approximation.
As to the contribution of the AVV three-point function (\ref{weak}), we find the following leading order expressions for the relevant components of $\Lambda^{\mu\nu\rho}(q_1,q_2)$ to the leading order in low momenta $q_1$ and $q_2$ ($|\bm{q}_{1,2}| \ll T$ and $|\omega_{1,2}| \ll T$)
\begin{align}
  \Lambda_{ij0}^{\text{tr}}(q_1,q_2) = \frac{16\ \pi T\,  \mathcal{C} \ \omega_2}{2 \pi T\, i\omega_2 - |\bm{q}_2|^2}\bigg[\epsilon_{ikj}q_{1k} + \frac{\epsilon_{klj}\, q_{2k}q_{1l} \, q_i}{2 \pi T \, i\omega - |\bm{q}|^2} \bigg]
  \label{J_spatial}
\end{align}
and
\begin{equation}
  \Lambda_{0j0}(q_1,q_2) = -32\, i \, \mathcal{C}\, \frac{\pi^2T^2 \omega_2}{(2\pi T \, i\omega - |\bm{q}|^2)(2\pi T \, i\omega_2- |\bm{q}_2|^2)}\, \epsilon_{jkl}q_{2k}q_{1l}  \;,
  \label{charge_0}
\end{equation}
with $q=q_1+q_2$ in Eqn. (\ref{kernel_cme}). The superscript ``tr'' of  $\Lambda_{ij0}^{\text{tr}}(q_1,q_2)$ refers to the component of $\Lambda_{ij0}(q_1,q_2)$ transverse to the momentum $\bm{q}_1$, i.e. $q_{1j}\Lambda_{ij0}^{\text{tr}}(q_1,q_2)=0$ and will be suppressed for the rest of the paper.  The diffusion denominators in (\ref{J_spatial}) and (\ref{charge_0}) imply a non-local response of the current to the chiral imbalance proxied by $\mathsf{A}_0(\omega_2, \bm{\mathfrak{q}}_2)$. In particular, we find the nontrivial infrared limits:
\begin{align}
  \lim_{\omega_2 \to 0}\lim_{\bm{q}_2\to 0}\Lambda_{ij0}(q_1,q_2) &= - 8 i \, \mathcal{C}\, \epsilon_{ikj}q_{1k} \;;
                                                                    \label{order_1} \\
  \lim_{\bm{q}_2\to 0} \lim_{\omega_2\to 0}\Lambda_{ij0}(q_1,q_2) &= 0 \;.
                                                                    \label{order_2}
\end{align}
Consequently, with $\mathsf{A}_0\sim\mu_A$, the CME signal for $|\bm{q}_2| \ll \sqrt{T |\omega_2|} \ll T$ and $|\omega_2| \ll |\bm{q}_2|^2/T \ll T$ can be quite different because of the AVV contribution. A simultaneous derivative expansion with respect to both $\mathsf{V}_\mu$ and $\mathsf{A}_0$ no longer exists. The CME current \eqref{eq:21} emerges when the momentum of $\mathsf{A}_0$, $q_2$, is close to the limit \eqref{order_2}. In the opposite situation when $q_2$ is close to \eqref{order_1}, the AVV contribution may enhance or reduce the CME signal depending on the sign of $\mathsf{A}_0$. In addition, a space-time dependent chiral imbalance induces a charge distribution via \eqref{charge_0}. The next order terms of $\Lambda_{ij0}$ and $\Lambda_{0j0}$ in low momenta have also been worked out.

From field theoretic perspectives\cite{Miklos, Hou, Feng}, the limit (\ref{order_1}) follows from the anomalous Ward identity which is robust because of the non-renormalization theorem \cite{Adler, Adler_Bardeen} of the chiral anomaly and the limit (\ref{order_2}) can be deduced from Coleman-Hill theorem \cite{Coleman_Hill}. The expression for arbitrary $q_1$ and $q_2$ is not robust against higher order corrections. The interpolating formulas (\ref{J_spatial}) and (\ref{charge_0}) pertain to the strong coupling, and are among the new outcomes of this work. Moreover, the underlying assumption of the Coleman-Hill theorem, the absence of infrared singularity in the zero momenta limit, is supported by our results.  In addition, we are able to prove UV and IR convergence of the kernel $\Lambda^{ij0}(q_1,q_2)$ and $\Lambda^{0j0}(q_1,q_2)$ in Eqn.~\eqref{weak}. The absence of UV divergence reflects the finiteness of the underlying dynamics of the $\mathcal{N}=4$ super-Yang-Mills.  

This paper is organized as follows. In the Section~\ref{sec:action}, we lay out the Einstein-Maxwell-Chern-Simons action in the bulk along with the equations of motion and link the solutions of EOM to the vector and axial-vector current on the boundary. The methodology of solving EOM analytically up to the required order is discussed in Section~\ref{sec:four-form-vect}. The main results on the chiral magnetic current in the presence of an constant axial chemical potential, a spacetime-dependent magnetic field and a spacetime-dependent $\mathsf{A}_0$ are presented in Section~\ref{sec:smallq} and the proof of UV/IR convergence is presented in Section~\ref{sec:uv-ir-convergence}. Section~\ref{sec:concluding} concludes the paper. Some technical details behind the solutions of EOM are deferred to Appendices \ref{sec:inhom-maxw-equat}-\ref{sec:diffusion}. For the benefit of readers, the frequently used notations that may cause confusions throughout the subsequent sections are tabulated in Appendix~\ref{sec:notation-list}.

% %%%%%%%%%%%%%%%%%%%%%%%%%%%%%%%%%%%%%%%%%

\section{The Einstein-Maxwell-Chern-Simons Action in Asymptotic $AdS_5$ Background}
\label{sec:action}

According to AdS/CFT correspondence \cite{Maldacena, Witten}, the $\mathcal{N}=4$ super-Yang-Mills theory at large $N_c$ (number of colors) and strong `t Hooft coupling in a 3+1 dimensional spacetime corresponds to the classical supergravity limit of type IIB superstring theory in an asymptotic AdS$_5$ spacetime with the 3+1 dimensions as its boundary. Consequently, the vector and axial-vector current correlators as well as the chiral anomalies of the super Yang-Mills can be described holographically with the following classical Einstein-Maxwell-Chern-Simons action in the asymptotic AdS$_5$ bulk \cite{Freedman, Yee2009, Gynther2011}, 
\begin{align}
  S = S_\text{EH} + S_{_\text{MCS}}  + S_\text{c.t.} \quad .
  \label{eq:1}
\end{align}
where $S_\text{EH}$ is the Einstein-Hilbert action
\begin{equation}
  S_\text{EH} = \kappa_{\text{EH}}\int \ \mathrm{d}^5X \sqrt{-g}\left(R-2\Lambda\right)
\end{equation}
with the curvature scalar $R$ and the negative cosmological constant $\Lambda = - \frac{12}{L^2}$, $S_{_\text{MCS}}$ is the Maxwell-Chern-Simons action and 
$S_\text{c.t.}$ is the holographic counter-terms (residing on the AdS boundary) to remove UV divergences of various holographic correlation functions.\footnote{The well-known holographic two-point function needs such counter-term to cut-off its UV divergence. In section~\ref{sec:uv-ir-convergence}, we will prove the finiteness of AVV correlation, hence counter-terms have nothing to do for CME three-point function. }  As will be shown in Section \ref{sec:uv-ir-convergence}, the AVV three-point function is free from UV divergence and thereby does not need counter terms. $S_\text{c.t.}$ is merely used to cancel the logarithmic divergence in two-point Green's functions and is not relevant to us. In terms of the left-hand and right-hand vector potentials $\mathrm{A}_L$ and $\mathrm{A}_R$, the Maxwell-Chern-Simons action reads
\begin{align}
  S_{_\text{MCS}} = & \kappa_{\text{M}}  \int \ \mathrm{d}^5X \sqrt{-g} \bigg[ -\frac{ 1 }{ 4 } (\mathrm{F}_L)_\text{MN} (\mathrm{F}_L)^\text{MN}  -\frac{ 1 }{ 4 } (\mathrm{F}_R)_\text{MN} (\mathrm{F}_R)^\text{MN} \nonumber\\
                                                                                                                                              &  + \frac{ \kappa_{\text{CS}} \epsilon^{MNOPQ} }{ 4\kappa_{\text{M}}\sqrt{-g} } \big( (\mathrm{A}_L)_M (\mathrm{F}_L)_{NO} (\mathrm{F}_L)_{PQ} - (\mathrm{A}_R)_M (\mathrm{F}_R)_{NO} (\mathrm{F}_R)_{PQ}\big) \bigg]+S_B \; .\label{eq:3}
\end{align}
where $M,N,O,P,Q$ refer to the indexes of the 4+1 dimensional bulk with the Levi-Civita symbol $\varepsilon^{MNOPQ}$ normalised according to $\varepsilon^{50123}=1$, the field strengths $(\mathrm{F}_L)_{MN}=\partial_M (\mathrm{A}_L)_N-\partial_N (\mathrm{A}_L)_M$, 
$(\mathrm{F}_L)_{MN}=\partial_M (\mathrm{A}_L)_N-\partial_N (\mathrm{A}_L)_M$, and $S_B$ is a boundary term to be specified later. The bulk part of this action is invariant under a 
$U_L(1)\times U_R(1)$ gauge transformation,
\begin{equation}
  (\mathrm{A}_L)_M\to (\mathrm{A}_L)_M+\partial_M(\phi_L)  \quad ,\quad  (\mathrm{A}_R)_M\to (\mathrm{A}_R)_M+\partial_M(\phi_R)  \;.
\end{equation}
Here, the gauge potentials $\mathrm{A}_L$ and $\mathrm{A}_R$ stem from the $U(1)$ subgroups of the global $U(4)$ symmetry of the $\mathcal{N}=4$ super-Yang-Mills theory on the boundary. Being tied to a global symmetry on the boundary, the $\mathrm{A}_L$ and $\mathrm{A}_R$ do not contribute the internal lines of the Feynman diagrams of the super Yang-Mills theory and is only employed in its gravity dual to generate various current correlations and thereby various transport coefficients of the super Yang-Mills plasma.

To describe the conserved vector current and anomalous axial vector current on the boundary, it is convenient to express the action in terms of the vector and axial vector gauge potentials via
\begin{align}
  \mathrm{A} = \frac{ 1 }{ \sqrt{2} } \left( \mathrm{A}_L - \mathrm{A}_R \right)  \quad ,\quad 
  \mathrm{V} = \frac{ 1 }{ \sqrt{2} }  \left( \mathrm{A}_L + \mathrm{A}_R \right) \;.
  \label{eq:4}
\end{align}
Integrating by part to remove $\mathrm{V}$ outside $\mathrm{F}^{\mathrm{V}}$ in the Chern-Simon's term and choose 
\begin{align}
  S_B = \int \ \mathrm{d}^5X \frac{\kappa_{\text{CS}}}{\sqrt{2}} \nabla_M \big[\varepsilon^{MNOPQ} \mathrm{A}_N \mathrm{V}_O \mathrm{F}^{\mathrm{V}}_{PQ}\big] \;.
  \label{eq:5}
\end{align}  
to cancel the boundary term incurred, we end up with 
\begin{align}
  S_\text{MCS} = \kappa_{\text{M}} \int \ \mathrm{d}^5 X \sqrt{-g} &\,  \bigg[ -\frac{1}{4}\mathrm{F}_\mathrm{V}^2 - \frac{1}{4} \mathrm{F}_A^2  \nonumber \\
  & + \frac{\kappa_{_\text{CS}}}{4 \sqrt{2}\kappa_{\text{M}}\sqrt{-g}} \varepsilon^{MNOPQ} \big( 3 \mathrm{A}_M\mathrm{F}_{NO}^\mathrm{V} \mathrm{F}_{PQ}^\mathrm{V} + \mathrm{A}_M \mathrm{F}_{NO}^A \mathrm{F}_{PQ}^A\big) \bigg] \; ,
  \label{eq:2}
\end{align}
The Maxwell-Chern-Simons action is invariant under an arbitrary $U_\mathrm{V}(1)$ gauge transformation
$\mathrm{V}_M\to \mathrm{V}_M+\partial_M\phi^\mathrm{V}$, but is invariant only under an axial $U_A(1)$ transformation $\mathrm{A}_M\to \mathrm{A}_M+\partial_M\phi^A$ with $\phi^A=0$ on the boundary.
Consequently, the strongly-coupled gauge theory on the boundary maintains only the $U_V(1)$ invariance. The $U_A(1)$ becomes anomalous with the parameters $\kappa_{\text{M}} $ and $\kappa_{_\text{CS}}$ determined by the anomaly coefficient. $S_B$ of (\ref{eq:5}) plays the role of the Bardeen term. The vector current associated to $U_V(1)$ is the analog of the electric current underlying CME with the corresponding charge referred to as the R-charge in the literature of super Yang-Mills.

The equations of motion corresponding to (\ref{eq:2}) can be readily obtained via variational principle and we have
\begin{align}
 \nabla_N \big[ (\mathrm{F}^{\mathrm{V}})^{NM} \sqrt{-g} \big] &= -  \frac{3}{2 \sqrt{2}}\frac{\kappa_{_\text{CS}}}{\kappa_{\text{M}}} \cdot \epsilon^{MNOPQ} (\mathrm{F}^{\mathrm{A}})_{NO} (\mathrm{F}^{\mathrm{V}})_{PQ} \label{eq:6} \\
   \nabla_N\big[ (\mathrm{F}^{\mathrm{A}})^{NM} \sqrt{-g} \big] &= -  \frac{3}{2 \sqrt{2}}\frac{\kappa_{_\text{CS}}}{\kappa_{\text{M}}}  \cdot \frac{1}{2} \epsilon^{MNOPQ} \bigg[ (\mathrm{F}^{\mathrm{V}})_{NO} (\mathrm{F}^{\mathrm{V}})_{PQ} + (\mathrm{F}^{\mathrm{A}})_{NO} (\mathrm{F}^{\mathrm{A}})_{PQ} \bigg] \;,\label{eq:7}
 \end{align}
 and
\begin{equation}
  R_{MN}-\frac{1}{2} R g_{MN}-\Lambda \, g_{MN} = -\frac{\kappa_{\text{M}}}{\kappa_{\text{EH}}}T_{MN} \;,\label{eq:1_ren} 
\end{equation}
with the scaled stress tensor:
\begin{equation}
\begin{aligned}
  T_{MN} &= \frac{2}{\kappa_{\text{M}}}\frac{\delta S_\text{MCS}}{\delta g^{MN}} \\
  &= (\mathrm{F}^{\mathrm{V}})_M^L(\mathrm{F}^{\mathrm{V}})_{NL} -\frac{1}{4}g_{MN}\, (\mathrm{F}^{\mathrm{V}})_{KL}(\mathrm{F}^{\mathrm{V}})^{KL} + (\mathrm{F}^{\mathrm{A}})_M^L(\mathrm{F}^{\mathrm{A}})_{NL} -\frac{1}{4}g_{MN}\, (\mathrm{F}^{\mathrm{A}})_{KL}(\mathrm{F}^{\mathrm{A}})^{KL}   \;.
\end{aligned}
\end{equation}
In the natural units, the mass dimensions of the coupling constants in the chiral action~\eqref{eq:2} are $\texttt{dim} \; \kappa_M =1$ and $\texttt{dim} \; \kappa_{_\text{CS}} =0$, hence we get the same mass dimension of gauge fields in $D=4+1$ dimensional spacetime as we have in $D=3+1$ dimensional QFT: $\texttt{dim}\; \mathrm{A}_M =\texttt{dim}\; \mathrm{V}_M = 1$ in coordinate representation.

It follows from the dictionary of AdS/CFT that the quantum effective action in the presence of vector and axial vector gauge potentials $\mathrm{V}_\mu$, $\mathrm{A}_\mu$, together with  3+1 dimensional metric $g_{\mu\nu}$ corresponds to the classical action (\ref{eq:1}) evaluated in terms of the solutions of EOM with the AdS-boundary values $\mathrm{V}_\mu$, $\mathrm{A}_\mu$ and $g_{\mu\nu}$ of respective fields. Taking the functional derivatives with respect to the boundary values of $\mathrm{V}_\mu$ and $\mathrm{A}_\mu$, we derive the holographic formulas of the vector and axial-vector currents:
\begin{align}
  J_\mathrm{V}^\mu(x) \equiv \frac{\delta S_{_\text{MCS}}}{\delta \mathrm{V}_\mu}\bigg|_{\text{AdS-boundary}}  &= \bigg[ -\kappa_{\text{M}} (\mathrm{F}_\mathrm{V})^{5 \mu} \sqrt{-g} + \frac{ 3 \, \kappa_{_\text{CS}} }{ \sqrt{2} } \, \epsilon^{\mu \nu \rho \sigma} \, {\mathrm{A}}_\nu (\mathrm{F}_\mathrm{V})_{\rho \sigma} \bigg]  \bigg|_{\text{AdS-boundary}}  \label{eq:8}\\
  J_{\mathrm{A}}^\mu(x) \equiv \frac{\delta S_{_\text{MCS}}}{\delta {\mathrm{A}}_\mu}\bigg|_{\text{AdS-boundary}}  &= \bigg[ -\kappa_{\text{M}} (\mathrm{F}_{\mathrm{A}})^{5 \mu} \sqrt{-g} + \frac{ \kappa_{_\text{CS} }}{ \sqrt{2} }\, \epsilon^{\mu \nu \rho \sigma} \, {\mathrm{A}}_\nu (\mathrm{F}_{\mathrm{A}})_{\rho \sigma}  \bigg]  \bigg|_{\text{AdS-boundary}}  \,  \label{eq:9}
\end{align}
where Greek indexes refer to the 3+1 dimensional spacetime on the boundary. It follows from the EOM (\ref{eq:6}) and (\ref{eq:7}) for $M=5$ that the vector current is conserved, while the divergence of the axial vector current acquires an anomaly, i.e.,
\begin{align}
  \partial_\mu J_{\mathrm{V}}^\mu(x) &=0  \label{eq:10}\\
  \partial_\mu J_{\mathrm{A}}^\mu(x) &= \frac{ \mathcal{C}}{ 3 } \epsilon^{\mu \nu \rho \sigma} \bigg[ 3 \, ( \mathrm{F}^{\mathrm{V}})_{\mu\nu} ( \mathrm{F}^{\mathrm{V}})_{\rho \sigma} +  ( \mathrm{F}^{\mathrm{A}})_{\mu\nu} ( \mathrm{F}^{\mathrm{A}})_{\rho \sigma} \bigg] \bigg|_{\text{AdS-boundary}} \;, \label{eq:11}
\end{align}
where the anomaly coefficient $\mathcal{C}$ is related to the Chern-Simons coupling via
\begin{equation}
  \mathcal{C} = \frac{3\kappa_{_\text{CS}}}{4\sqrt{2}} \;.
  \label{anomaly_coeff}
\end{equation}

 We notice that for the $\mathcal{N}=4$ $SU(N_c)$ super Yang-Mills at large $N_c$ and strong `t Hooft coupling, both $\kappa_{\text{EM}}$ and $\kappa_{\text{M}}$ scales with $N_c^2$ as $N_c\to\infty$ \cite{Freedman, Policastro2002a}, hence the coefficient on RHS of (\ref{eq:1_ren}) for the strength of the gravitational coupling, $\kappa_{\text{M}}/\kappa_{\text{EH}} = \mathcal{O}(1)$ and is thereby not tunable. 
Introducing the fluctuations $\{h_{\mu\nu}, \mathbb{A}_\mu, \mathbb{V}_\mu \}$ from the background $\{\bar{g}_{\mu\nu}, \bar{A}_\mu, \bar{V}_\mu \}$ via
\begin{align}
  g_{\mu\nu} = \bar g_{\mu\nu} + h_{\mu\nu} \; , \; \mathrm{A}_{\mu} = \bar{\mathrm{A}}_{\mu} + \mathbb{A}_\mu \; , \; \mathrm{V}_{\mu} = \bar{\mathrm{V}}_{\mu} + \mathbb{V}_\mu \;, \label{eq:60}
\end{align}
EOM (\ref{eq:6}), (\ref{eq:7}) and (\ref{eq:1_ren}) become a set of nonlinear equations for the fluctuations and we shall impose the radial gauge condition $V_5 = {\mathrm{A}}_5 = g_{5 \star}=0$ for their solution.  Substituting the solution into (\ref{eq:8}) and (\ref{eq:9}), we obtain the vector and axial-vector currents as functionals of the boundary values of the fluctuating fields, which is the holographic version of (\ref{weak}) and can be expanded in powers of the AdS-boundary values.

To explore the chiral magnetic effect at a nonzero temperature and zero R-charge chemical potentials in strong coupling, we start with EoM with the background solution of the AdS-Schwarzschild metric and zero gauge potentials, i.e.
\begin{align}
  \mathrm{d} s^2_5 = \bar g_{MN}dx^Mdx^N = \frac{ (\pi L T)^2 }{ u } \left( -f(u)\, \mathrm{d} t^2 + \sum\limits_{i=1}^3 \mathrm{d} (x^i)^2 \right) +\frac{ 1  }{ 4u^2 f(u) } \, \mathrm{d} u^2 \; , f(u) = 1-u^2 \;,
  \label{eq:16}
\end{align}
and
\begin{equation}
  \bar{\mathrm{V}}_M = \bar{\mathrm{A}}_M =0,
\end{equation}
where $u=0$ is the AdS-boundary and $L$ the AdS radius. The Hawking temperature $T$ of the horizon $u=1$ corresponds to the temperature of thermal bath of the boundary field theory. It is interesting to notice the following power structure of 
the respective equations of (\ref{eq:6})-(\ref{eq:1_ren}) in this AdS-Schwarzschild geometry:
\begin{align}
  &\text{Vector gauge field}:& \quad \mathcal{O}(\mathbb{V}) + \mathcal{O}(h \mathbb{V}) &= \mathcal{O}(\mathbb{{\mathbb{A}} V}) & \label{eq:14} \\
  &\text{Axial gauge field}:& \quad \mathcal{O}(\mathbb{A}) + \mathcal{O}(h \mathbb{A}) &= \mathcal{O}(\mathbb{A}^2) + \mathcal{O}(\mathbb{V}^2)  & \label{eq:15} \\
  &\text{Metric field}:& \quad  \mathcal{O}(h) &= \mathcal{O}(\mathbb{V}^2) + \mathcal{O}(\mathbb{A}^2)  \;. & \label{eq:16_ren} 
\end{align}
It follows that, because of Eqn.~\eqref{eq:16_ren}, the terms in Eqns.~\eqref{eq:14} and \eqref{eq:15} involving the metric fluctuations entail the cubic fluctuations of gauge fields, therefore do not contribute to the quadratic order of Eqns.(\ref{eq:6}) and (\ref{eq:7}) in gauge field fluctuations and thereby do not contribute to the critical term of the chiral magnetic current, i.e. $\mathcal{O}(\mathbb{A V})$ term of the vector current (\ref{eq:8}). Specifically, all we need to do is to solve the Maxwell-Chern-Simons equations (\ref{eq:6}) and (\ref{eq:7}) with $(V, A)$ replaced by the fluctuations $\{\mathbb{A},\mathbb{V}\}$ and the metric fixed to the AdS-Schwarzschild background (\ref{eq:16}),  i.e. 
\begin{align}
  \partial_N \big[\bar g^{NP}\bar g^{MQ}(\mathbb{F}^V)_{PQ} \sqrt{-\bar g} \big] &= -  \frac{3}{2 \sqrt{2}}\frac{\kappa_{_\text{CS}}}{\kappa_{\text{M}}} \cdot \epsilon^{MNOPQ} (\mathbb{F}^A)_{NO} (\mathbb{F}^V)_{PQ} \label{eq:6_ren}  \\
  \partial_N \big[\bar g^{NP}\bar g^{MQ}(\mathbb{F}^A)_{PQ} \sqrt{-\bar g} \big] &= -  \frac{3}{2 \sqrt{2}}\frac{\kappa_{_\text{CS}}}{\kappa_{\text{M}}}  \cdot \frac{1}{2} \epsilon^{MNOPQ} \bigg[ (\mathbb{F}^V)_{NO} (\mathbb{F}^V)_{PQ} + (\mathbb{F}^A)_{NO} (\mathbb{F}^A)_{PQ} \bigg] \;,\label{eq:7_ren}.
\end{align}
The AdS-boundary conditions $\mathbb{V}_0(x^\mu;u)\big|_{u \to 0} \equiv 0$ because of the zero R-charge chemical potential we assumed; $\vec{\mathbb{V}}(x^\mu;u)\big|_{u \to 0} \ne 0$ that leads to a spacetime-dependent magnetic field $\bf{\mathbb{B}(x^\mu)}$; and $\mathbb{A}_0(x^\mu;u)\big|_{u \to 0} \ne 0$ that proxies a spacetime-dependent chiral imbalance in a strongly-coupled quark-gluon plasma. Substituting the solution to the currents (\ref{eq:8}) and (\ref{eq:9}), the term of the spatial component of the vector current that is bilinear in $\bm{\mathsf{B}}(x^\mu)$ and $\mathsf{A}_0(x^\mu)$ gives rise to CME. Its coefficient corresponding to the $(0ij)$ component of the AVV triangle diagram in field theory includes all orders of the $\mathcal{N}=4$ super-Yang-Mills coupling and is to be evaluated analytically in this work. The Maxwell-Chern-Simons equations (\ref{eq:6_ren}) and (\ref{eq:7_ren}) become a set of coupled nonlinear equations with respect to the 
fluctuating fields and can be solved iteratively. The order of magnitude sorting described in (\ref{eq:14}), (\ref{eq:15}) and (\ref{eq:16_ren}) applies to weak magnetic field and chiral imbalance in a thermal bath of high temperature. For the quark-gluon plasma created in RHIC, the temperature is $200-300 \,\mathrm{MeV}$ and the magnetic field is of the order of $\sqrt{m_\pi}$ and $A_0$ is difficult to estimate. The weak field approximation appears marginal.

In contrast, an alternative background geometric that corresponds to a nonzero temperature and a nonzero chemical potential is the AdS-Reissner–Nordström (AdS-RN) geometry, which is accompanied by a nonzero background gauge potential, $\mathrm{V}_0\neq 0$. The power structure of the Einstein-Maxwell-Chern-Simons equations become
\begin{align}
  &\text{Vector gauge field}:& \quad \mathcal{O}(\mathbb{V}) + \mathcal{O}(h) + \mathcal{O}(h \mathbb{V}) &= \mathcal{O}(\mathbb{A}) + \mathcal{O}(\mathbb{A V})& \label{eq:14_ren} \\
  &\text{Axial gauge field}:& \quad \mathcal{O}(\mathbb{A}) + \mathcal{O}(h \mathbb{A}) &= \mathcal{O}(h) + \mathcal{O}(\mathbb{A}^2) + \mathcal{O}(\mathbb{V}^2)&   \label{eq:15_ren} \\
  &\text{Metric field}:& \quad  \mathcal{O}(h) &= \mathcal{O}(\mathbb{V}) + \mathcal{O}(\mathbb{V}^2) + \mathcal{O}(\mathbb{A}^2)   \;. &   \label{eq:17_ren} 
\end{align}
Consequently, once the background chemical potential is introduced, the metric fluctuations can't be decoupled even at linear order of fluctuation in the vector and axial vector gauge fields, which complicates the analytic calculations for the three-point functions $\Delta_{\rho \mu \nu}$.

On the other hand, the expression of the AVV three point function $\Lambda_{ij0}(q_1,q_2)$ with $q_2=(\omega_2,0)$ can be deduced from the anomalous Ward identity (\ref{eq:11}) and is thereby robust to all orders of metric fluctuations in the presence of a AdS-RN black hole. The momentum representation of (\ref{eq:11}) implies that
\begin{equation}
  i\, (k_1+k_2)_\mu\Delta^{\mu\rho\lambda}(k_1,k_2) = 8\, \mathcal{C}\, \epsilon^{\mu\nu\rho\lambda} k_{1 \mu} k_{2 \mu}
\end{equation}
with $k_1$ and $k_2$ the 4-momenta of the boundary value of the vector gauge potential (one associated to the magnetic field and the other to the vector current). It follows from (\ref{eq:34}) that 
\begin{equation}
  i\, q_{2\mu}\Lambda^{\mu\rho\lambda}(q_1,q_2) = 8\, \mathcal{C}\, \epsilon^{\mu\nu\rho\lambda} q_{2 \mu}q_{1 \nu}
\end{equation}
For the special $q_2$ assumed above, it follows readily that
\begin{equation}
  \Lambda_{ij0}(q_1,q_2) = - 8i\, \mathcal{C}\, \epsilon_{ikj}q_{1 k}
  \label{asymptotic}
\end{equation}
Though the constraint (\ref{eq:55}) is imposed for the axial vector potential on the boundary, eq. (\ref{asymptotic}) serves an asymptotic form of $J_{AVV}^\mu$ for a nearly homogeneous $\mathsf{A}_0$ there.

At this point, we would like to clarify the relationship between our weak field approximation and the hydrodynamics approximation under external vector and axial vector gauge potentials. In the framework of the power counting presented above, the utilization of hydrodynamic approximation demands the condition $|\mathsf{A}_0(q)|  \ll \mu_A$ that amounts to resum all powers of $\mu_A$, leaving $\mathsf{A}_0(q)$ and the magnetic field as perturbations. As we shall see in eq.~\eqref{eq:57} below, $\mathsf{A}_0(q)$ and $\mu_A$ correspond to the values of $\mathcal{A}_0(q|u)$ on the AdS-boundary and at the horizon, respectively. Consequently, a resummation of $\mu_A$ involves all powers of $\mathcal{A}_0$ and those metric fluctuations cannot be avoided. Employing the probe approximation can dismiss the backreaction \cite{Gynther2011, Bu2016a}, and work on the AdS-Schwarzschild background, but the condition ${\kappa_{\text{M}}}/{\kappa_{\text{EM}}}\ll 1$ is  artificially required, which renders the boundary field theory not the super Yang-Mills.

% %%%%%%%% 
\section{The Solution Algorithm}
\label{sec:four-form-vect}

For the chiral magnetic effect under an arbitrarily space-time dependent magnetic field and chiral imbalance, we need the electric current (\ref{eq:8}) in terms of the boundary values.
\begin{equation}
  \mathbb{V}_\mu(x;0)=(0,\bm{\mathsf{V}}(x)) \;;\qquad \mathbb{A}_\mu(x;0)=(\mathsf{A}_0(x),\bm{0})
  \label{eq:boundary}
\end{equation}
where $\bm{\nabla}\cdot\bm{\mathsf{V}}=0$ and we have adapted the radial gauge condition $\mathbb{V}_u=\mathbb{A}_u=0$.  
The second term on RHS of (\ref{eq:8}) is already explicit in terms of (\ref{eq:boundary}). The explicit expression of the first term on RHS of (\ref{eq:8}) will be derived in this section. In what follows, we shall solve the nonlinear Maxwell-Chern-Simons equation (\ref{eq:6_ren}) and (\ref{eq:7_ren}) iteratively to find out $\mathbb{F}_{5\mu}=\frac{\partial\mathbb{V}_\mu}{\partial u}$ in terms of the boundary values and one iteration serves our purpose.

Using Chern-Simons coupling $\kappa_{_\text{CS}}$ to track the order of iteration, we have 
\begin{align}
  \mathbb{V}=\mathcal{V}  + \mathcal{O}(\kappa_{_\text{CS}}) =  \mathtt{V} +\mathcal{O}(\kappa_{_\text{CS}}^2) \label{eq:53_1}
\end{align}
and 
\begin{align}
  \mathbb{A}=\mathcal{A} + \mathcal{O}(\kappa_{_\text{CS}}) = \mathtt{A} +\mathcal{O}(\kappa_{_\text{CS}}^2)\label{eq:53_2}
\end{align}
where the zeroth order solutions $\mathcal{V}$ and $\mathcal{A}$ solve the linear homogeneous equations:
\begin{align}
  \partial_N \big[\bar g^{NP}\bar g^{MQ}(\mathcal{F}^V)_{PQ} \sqrt{-\bar g} \big] &=   0  \;;\label{eq:6_ren1}  \\
  \partial_N \big[\bar g^{NP}\bar g^{MQ}(\mathcal{F}^A)_{PQ} \sqrt{-\bar g} \big] &=   0  \;,\label{eq:7_ren1}
\end{align}
and the first iteration gives rise to $\mathtt{V}$ and $\mathtt{A}$ in (\ref{eq:53_1}) and (\ref{eq:53_2}). The differential equations satisfied by  $\mathtt{V}$ and $\mathtt{A}$ are obtained by replacing  $\mathbb{V}$ and $\mathbb{A}$ on RHS of (\ref{eq:6_ren}) and (\ref{eq:7_ren}) by the zeroth order solutions $\mathcal{V}$ and $\mathcal{A}$, i. e.
\begin{align}
  \partial_N \big[\bar g^{NP}\bar g^{MQ}(\mathtt{F}^V)_{PQ} \sqrt{-\bar{g}} \big] &= -  \frac{3}{2 \sqrt{2}} \frac{\kappa_{_\text{CS}}}{\kappa_{\text{M}}} \cdot \epsilon^{MNOPQ} (\mathcal{F}^A)_{NO} (\mathcal{F}^V)_{PQ} \;,\label{eq:6_ren2} \\
  \partial_N \big[\bar{g}^{NP} \bar{g}^{MQ}(\mathtt{F}^A)_{PQ} \sqrt{-\bar{g}} \big] &= -  \frac{3}{2 \sqrt{2}}\frac{\kappa_{_\text{CS}}}{\kappa_{\text{M}}}  \cdot \frac{1}{2} \epsilon^{MNOPQ} \bigg[ (\mathcal{F}^V)_{NO} (\mathcal{F}^V)_{PQ} + (\mathcal{F}^A)_{NO} (\mathcal{F}^A)_{PQ} \bigg] \;, \label{eq:7_ren2}
\end{align}
with $\mathcal{F}^V_{MN}= \partial_M \mathcal{V}_N - \partial_N \mathcal{V}_M$ and $\mathcal{F}^A_{MN}= \partial_M \mathcal{A}_N - \partial_N \mathcal{A}_M$, and are linear partial differential equations with inhomogeneous terms.

Because of the translation invariance with respect to boundary coordinates $x^\mu$, it is convenient to introduce the momentum representation via
\begin{align}
  \mathbb{V}_\mu(x;u) = \int \ \frac{ \mathrm{d}^4q}{(2 \pi)^4}e^{\mathrm{i}\, qx}\; \mathbb{V}_\mu(q|u) \;,
  \qquad 
  \mathbb{A}_\mu(x;u) = \int \ \frac{ \mathrm{d}^4q}{(2\pi)^4}e^{\mathrm{i}\, qx}\; \mathbb{A}_\mu(q|u) \;,
  \label{eq:17}
\end{align}
and the boundary condition (\ref{eq:boundary}) becomes
\begin{equation}
  \mathbb{V}_\mu(q|0)=(0,\bm{\mathsf{V}}(q)) \qquad \mathbb{A}_\mu(q|0)=(\mathsf{A}_0(q),\bm{0})
  \label{eq:boundary1}
\end{equation}
with $\bm{q}\cdot\mathsf{V}(q)=0$ ($\bm{\nabla}\cdot\bm{\mathsf{V}}=0$ in coordinate representation). The details of the Fourier transformation to the momentum representation are described in Appendix \ref{sec:inhom-maxw-equat}.

\subsection{Zeroth Order}
\label{sec:zeroth-order}

Carrying out the Fourier transformation prescribed in Appendix \ref{sec:inhom-maxw-equat} for $\{\mathcal{V},\mathcal{A}\}$, the leading order equations (\ref{eq:6_ren1}) and (\ref{eq:7_ren1}) reduce to 
\begin{align}
  \mathfrak{w} \mathcal{A}_0' + f\, \big( \bm{\mathfrak{q}} \cdot \bm{\mathcal{A}}\big)' &= 0   \;; \label{eq:19}\\
  \mathfrak{w} \mathcal{V}_0' + f\, \big(\bm{\mathfrak{q}} \cdot \bm{\mathcal{V}}\big)' &= 0  \;; \label{eq:20_0}
\end{align}
\begin{align}
  \mathcal{A}_0'' - \frac{1}{uf} \big[\bm{\mathfrak{q}}^2 \mathcal{A}_0 + \mathfrak{w}\, (\bm{\mathfrak{q}} \cdot \bm{\mathcal{A}}) \big] &= 0 \;; \label{eq:21_0}\\
  \mathcal{V}_0'' - \frac{1}{uf} \big[\bm{\mathfrak{q}}^2 \mathcal{V}_0 + \mathfrak{w}\, (\bm{\mathfrak{q}} \cdot \bm{\mathcal{V}})\big] &= 0  \;; \label{eq:22_0}
\end{align}
\begin{align}
  \mathcal{A}_k'' + \frac{f'}{f} \mathcal{A}_k' + \frac{1}{u f^2} \big[\mathfrak{w}^2 \mathcal{A}_k + \mathfrak{w} \mathfrak{q}_k\, \mathcal{A}_0\big] - \frac{1}{u f} \big[|\bm{\mathfrak{q}}|^2 \mathcal{A}_k - \mathfrak{q}_k\, (\bm{\mathfrak{q}} \cdot \bm{\mathcal{A}})\big] &= 0   \;;\label{eq:23}  \\
  \mathcal{V}_k'' + \frac{f'}{f} \mathcal{V}_k' + \frac{1}{uf^2} \big[\mathfrak{w}^2 \mathcal{V}_k + \mathfrak{w} \mathfrak{q}_k\, \mathcal{V}_0\big] - \frac{1}{u f} \big[|\bm{\mathfrak{q}}|^2 \mathcal{V}_k - \mathfrak{q}_k\, (\bm{\mathfrak{q}} \cdot \bm{\mathcal{V}})\big] &= 0  \;, \label{eq:24}
\end{align} %
where $f$ is the metric function $f \equiv 1-u^2 $ in \eqref{eq:16} and  we have defined the dimensionless momenta $(\mathfrak{w}, \bm{\mathfrak{q}})$:
\begin{align}
  \mathfrak{w} = \frac{\omega}{2\pi T} \;; \quad \bm{\mathfrak{q}} = \frac{q}{2 \pi T} \;.\label{eq:dimensionless}
\end{align}

Decomposing $\bm{\mathcal{V}}$ and $\bm{\mathcal{A}}$ into their transverse and longitudinal components with respect to the spatial momentum $\bm{q}$,
\begin{align}
  \bm{\mathcal{V}}_i(q|u) = (\bm{\mathcal{V}}_\perp)_i + (\bm{\mathcal{V}}_\parallel)_i &= (\delta_{ij} - \hat{\mathfrak{q}}_i \hat{\mathfrak{q}}_j)\bm{\mathcal{V}}_j(q|u) + \hat{\mathfrak{q}}_i \hat{\mathfrak{q}}_j \, \bm{\mathcal{V}}_j(q|u) \;,
                                                                                          \label{eq:36_0} \\  %%changed
  \bm{\mathcal{A}}_i(p|u) = (\bm{\mathcal{A}}_\perp)_i + (\bm{\mathcal{A}}_\parallel)_i &= (\delta_{ij} - \hat{\mathfrak{q}}_i \hat{\mathfrak{q}}_j)\, \bm{\mathcal{A}}_j(p|u) + \hat{\mathfrak{q}}_i \hat{\mathfrak{q}}_j \, \bm{\mathcal{A}}_j(p|u) \;,
                                                                                          \label{eq:37}
\end{align}
with the indices $ i, j = 1,2,3$ and $\hat{\mathfrak{q}}$ the unit vector in the direction of $\bm{\mathfrak{q}}$, 
we find that each component of $\bm{\mathcal{V}}_\perp$ and $\bm{\mathcal{A}}_\perp$ satisfies the following second order linear ordinary differential equation
\begin{equation}
  \bm{\varPsi}'' - \frac{2u}{1-u^2} \, \bm{\varPsi}' + \frac{\mathfrak{w}^2 - |\bm{\mathfrak{q}}|^2 (1-u^2)}{u(1-u^2)^2} \, \bm{\varPsi} = 0 \;, \label{eq:29}
\end{equation}
where $\bm{\varPsi}(q|u)= \{ \bm{\mathcal{V}}_\perp \;,\, \bm{\mathcal{A}}_\perp \}$ and we have substituted the explicit form of $f$. The temporal components $\mathcal{V}_0', \mathcal{A}_0'$ in eqs.~\eqref{eq:19} and \eqref{eq:20_0} can be decoupled from the longitudinal components by eliminating $(\bm{\mathfrak{q}}\cdot\bm{\mathcal{V}})$ and $(\bm{\mathfrak{q}}\cdot \bm{\mathcal{A}})$ from \eqref{eq:21_0} and \eqref{eq:22_0} and we end up with another second order linear ordinary differential equation
\begin{equation}
  \varPhi'' + \frac{1-3u^2}{u(1-u^2)} \varPhi' + \frac{\mathfrak{w}^2 - |\bm{\mathfrak{q}}|^2 (1-u^2)}{u(1-u^2)^2} \varPhi = 0 \;, \label{eq:36}
\end{equation}
for $\varPhi(q|u)=\{\mathcal{V}_0' , \mathcal{A}_0'\}$. The boundary conditions of. \eqref{eq:29} and \eqref{eq:36} follow from (\ref{eq:boundary1}) to this order and read
\begin{align}
  \bm{\mathcal{V}}_\perp(q|0) = \bm{\mathsf{V}}(q) \;; \quad  \lim_{u\to 0} u \mathcal{V}_0''(q|u) = 0 \;; \quad 
  \bm{\mathcal{A}}_\perp(q|0) = 0 \;; \quad   \lim_{u\to 0} u \mathcal{A}_0''(q|u) = |\bm{\mathfrak{q}}|^2\mathsf{A}_0(q) \;, \label{eq:36_ren}
\end{align}
where we have converted the boundary conditions for $\mathcal{V}_0$ and  $\mathcal{A}_0$ to that for  $\mathcal{V}_0''$ and  $\mathcal{A}_0''$ via eqs.(\ref{eq:21_0}) and (\ref{eq:22_0}). Since the equation for $\bm{\mathcal{A}}_\perp$ stands by itself, the zero boundary value $\bm{\mathcal{A}}_\perp$ implies the null solution $\bm{\mathcal{A}}_\perp=0$. 

Both of eqs.~\eqref{eq:29} and \eqref{eq:36} are the Fuchs equations with four regular points, and can be transformed into the standard Heun equation with details shown in Appendix~\ref{sec:heun}. The asymptotic behaviors near the boundary and near the horizon are exhibited in Tab.\ref{tab:indexes}.

\begin{table}[ht]
  \caption{$\Delta$ Indexes of power series and asymptotic behavior} % title of Table
  \setlength{\tabcolsep}{4mm}{ %% \setlength command to broaden the gap between entries in table
  \centering % used for centering table
  \begin{tabular}{c c c c c} % centered columns (4 columns)
    \hline\hline %inserts double horizontal lines
    Fluctuation &  Horizon &  AdS-boundary &   $u \to 1^-$ & $u \to 0$ \\ [0.5ex] % inserts table
    % heading
    \hline % inserts single horizontal line
    \vspace{.1cm} \\
    $\varPhi(q|u)$ & $\Delta_{_\text{H}} = \pm  \frac{\mathfrak{w}}{2}\mathrm{i}$   & $\Delta_{_\text{AdS}} =0; 0$ & $(1-u)^{\pm i\frac{\mathfrak{w}}{2}}$ & $\mathcal{O}(1)$, $\mathcal{O}\big( \log u \big)$ \vspace{.5cm} \\
    $\bm{\mathcal{V}}_\perp(q|u)$ & $\Delta_{_\text{H}} = \pm \frac{\mathfrak{w}}{2}\mathrm{i}$   & $\Delta_{_\text{AdS}} = 0; 1$ & $(1-u)^{\pm i\frac{\mathfrak{w}}{2}}$ & $\mathcal{O}(u)$, $\mathcal{O}\big(u\log u \big) + \mathcal{O}(1)$ \\ [1ex] % [1ex] adds vertical space
    \hline %inserts single line
  \end{tabular} }
  \label{tab:indexes}
\end{table}
For a retarded response, we choose the in-falling wave solution at horizon \cite{Policastro2002a}, which means $\Delta_{_\text{H}} = - \frac{\mathfrak{w}}{2} i$, and we have:
\begin{equation}
  \bm{\mathcal{V}}_\perp \propto (1-u)^{-i\frac{\mathfrak{w}}{2}} \;;\quad \mathcal{V}_0'\propto (1-u)^{-i\frac{\mathfrak{w}}{2}} \;;\quad \mathcal{A}_0'\propto (1-u)^{-i\frac{\mathfrak{w}}{2}} \;, \quad \text{as} \; u \to 1^- \;. \label{eq:13}
\end{equation}
Denoting the pairs of linearly independent solutions of (\ref{eq:29}) and (\ref{eq:36}) by $\{\psi_1(q|u), \psi_2(q|u)\}$ and $\{\phi_1(q|u), \phi_2(q|u)\}$, respectively, we identify $\psi_1(q|u)$ and $\phi_1(q|u)$ as the in-falling wave solutions of (\ref{eq:29}) and (\ref{eq:36}), normalized at the horizon according to \footnote{The other index $\Delta_\mathrm{H} = + \frac{\mathfrak{w}}{2}\mathrm{i}$ assigned to $\psi_2(q|u)$ and $\phi_2(q|u)$ leads to the advanced response.}
\begin{align}
  \lim_{u\to 1^-}(1-u)^{ i\frac{\mathfrak{w}}{2}}\psi_1(q|u)=1 \;;\qquad 
  \lim_{u\to 1^-}(1-u)^{i\frac{\mathfrak{w}}{2}}\phi_1(q|u)=1 \;.
  \label{normalization}
\end{align}
As shown in Appendix~\ref{sec:heun}, $\psi_1(q|u)$ and $\phi_1(q|u)$ can be expressed in terms of two Heun functions. Following the boundary condition ~\eqref{eq:36_ren}, the solutions of the zeroth order take the forms 
\begin{equation}
  \begin{aligned}
    \bm{\mathcal{V}}(q|u) &= \bm{\mathsf{V}}(q)\, \frac{\psi_1(q|u) }{\psi_1(q|0)} \;; &\quad& \mathcal{V}_0(q|u) = 0  \;; \\
    \mathcal{A}_0'(q|u) &= |\bm{\mathfrak{q}}|^2\mathsf{A}_0(q) \,\frac{\phi_1(q|u)}{D_1(q)}  \;; &\quad& \bm{\mathcal{A}}_\perp(q|u) = 0  \;;
  \end{aligned}
  \label{free}
\end{equation}
with  
\begin{equation}
  D_1(q)=\lim_{u\to 0}u\phi_1'(q|u) \;.\label{eq:38}
\end{equation} 

The solution of $\mathcal{A}_0(q|u)$ at $\bm{q}=0$ is related to the axial chemical potential and requires special treatment in order to be conjugate to a macroscopic axial charge density. Let us return to the coordinate representation and seek a solution of $\mathcal{A}_0(X)$ that is independent the spatial coordinates on the boundary. It follows readily from eq.\eqref{eq:7_ren1} that such a $\mathcal{A}_0$ satisfies the equation 
\begin{equation}
  \partial_0\partial_5\mathcal{A}_0=0  \;;\qquad \partial_5^2\mathcal{A}_0=0 \;,
\end{equation}
which implies the solution $\mathcal{A}_0=au+b$, where $a$ is time-independent and $b$ can depend on time. As only $(\mathcal{F}^A)_{MN}$ contributes to RHS of (\ref{eq:6_ren2}) and (\ref{eq:7_ren2}), the first term of the current (\ref{eq:8}) does not depend $b$ but the second term does. Following the gauge invariant definition of the axial chemical potential proposed in \cite{Gynther2011,Rubakov2010}, we have
\begin{equation}
  \mathcal{A}_0(X) \bigg|_{\bm{q}=0} = \mu_A u \;, \label{eq:57_0}
\end{equation}
whose 4-momentum representation reads
\begin{equation}
  \mathcal{A}_0(q|u) \bigg|_{\bm{q}=0} = (2\pi)^4\delta^4(q)\mu_A u \;, \label{eq:57}
\end{equation}
Consequently  $\mathcal{A}_0(q|0) \bigg|_{\bm{q}=0}\equiv \mathsf{A}_0(q) \bigg|_{\bm{q}=0} =0$.

\subsection{First Order}
\label{sec:first-order}

To calculate the chiral magnetic current, we need only to carry out the iteration to the first order for the vector gauge potential $\mathbb{V}_\mu$. Substituting the zeroth order solution (\ref{free}) into RHS of eq.~\eqref{eq:6_ren2} and making Fourier transformation with respect to the boundary coordinates $x$, we obtain that 
\begin{equation}
  \mathfrak{w} \mathtt{V}_0' + f\, \big(\bm{\mathfrak{q}} \cdot \bm{\mathtt{V}}\big)' = \frac{\kappa_{_\text{CS}}}{\kappa_{\text{M}}} G^5_{_\mathtt{V}}(q|u)  \label{eq:20}
\end{equation}
\begin{equation}
  \mathtt{V}_0'' - \frac{1}{u f} \big[|\bm{\mathfrak{q}}|^2 \mathtt{V}_0 + \mathfrak{w}\, (\bm{\mathfrak{q}} \cdot \bm{\mathtt{V}})\big] = \frac{\kappa_{_\text{CS}}}{\kappa_{\text{M}}} G^0_{_\mathtt{V}}(q|u)  \label{eq:22}
\end{equation}
\begin{equation}
  \mathtt{V}_k'' + \frac{f'}{f} \mathtt{V}_k' + \frac{1}{u f^2} \big[\mathfrak{w}^2 \mathtt{V}_k + \mathfrak{w} \mathfrak{q}_k\, \mathtt{V}_0\big] - \frac{1}{u f} \big[|\bm{\mathfrak{q}}|^2 \mathtt{V}_k - \mathfrak{q}_k\, (\bm{\mathfrak{q}} \cdot \bm{\mathtt{V}})\big] = \frac{\kappa_{_\text{CS}}}{\kappa_{\text{M}}}G_{_\mathtt{V}}^k(q|u)  \;.  \label{eq:24}
\end{equation}
Here we need to distinguish $\bm{\mathcal{A}}\bigg|_{\bm{q}=0}$ from $\bm{\mathcal{A}}\bigg|_{\bm{q}\ne 0}$. In case of the former, it follows from the discussion towards the end of the last sub-section that
\begin{equation}
  \bm{G}_{_\mathtt{V}}(q|u) = - \frac{3}{\sqrt{2}} \frac{\mu_A}{(\pi T)^2 L f} \bm{\mathcal{B}}(q|u) \;;\qquad  G_{_\mathtt{V}}^0(q|u) = G_{_\mathtt{V}}^5(q|u) = 0 \;.
  \label{G_mu_A}
\end{equation}
In case of the latter, each component of $\bm{G}_{_\mathtt{V}}^{M}(q|u)$ is a convolution of the zeroth order solution of $\mathcal{V}$ and  $\mathcal{A}$, i.e.
\begin{align}
  G_{_\mathtt{V}}^M(q|u) = \int \frac{\mathrm{d}^4q_1}{(2\pi)^4} \frac{\mathrm{d}^4q_2}{(2\pi)^4} (2\pi)^4 \delta^4(q_1+q_2-q)\; \mathcal{G}_{_\mathrm{V}}^M(q_1,q_2|u)   \;,
  \label{eq:25}
\end{align}
where the integrands read 
\begin{align}
  \mathcal{G}_{_\mathrm{V}}^5(q_1,q_2|u) &= -\frac{3}{\sqrt{2}} \frac{u f}{(\pi T)^2L} \frac{1}{|\bm{\mathfrak{q}}_2|^2}\mathcal{A}_0''(q_2 |u)\; \big( \bm{\mathfrak{q}}_2 \cdot  \bm{\mathcal{B}}(q_1|u) \big)  \label{eq:31} \;;  \\
  \mathcal{G}_{_\mathrm{V}}^0(q_1,q_2|u) &=  \frac{3}{\sqrt{2}} \frac{1}{(\pi T)^2 L\, f}\frac{\mathfrak{w}_2}{|\bm{\mathfrak{q}}_2|^2} \mathcal{A}_0'(q_2|u)\; \big( \bm{\mathfrak{q}}_2 \cdot \bm{\mathcal{B}}(q_1|u) \big)  \label{eq:33}  \;;   \\
  \bm{\mathcal{G}}^V(q_1,q_2|u) &= - \frac{3}{\sqrt{2}} \frac{1}{(\pi T)^2 L f} \bigg[ \mathcal{A}_0'(q_2|u) \bm{\mathcal{B}}(q_1|u) \nonumber \\
  & \qquad - \frac{\mathfrak{w}_2}{f\, |\bm{\mathfrak{q}}_2|^2} \mathcal{A}_0'(q_2|u)\; \bm{\mathfrak{q}}_2 \times \bm{\mathcal{E}}(q_1|u) - \mathrm{i}\, u f \frac{2 \pi T}{{|\bm{\mathfrak{q}}_2|^2}} \mathcal{A}_0''(q_2|u) \; \bm{\mathfrak{q}}_2 \times \mathcal{V}'(q_1|u) \bigg]  \label{eq:35} \;.
\end{align}
The spatial vectors $\bm{\mathcal{B}}(q_1|u)$ and $\bm{\mathcal{E}}(q_1|u)$ in eq.~\eqref{G_mu_A}\eqref{eq:31}\eqref{eq:33}\eqref{eq:35} are related to the magnetic field $\bm{\mathsf{B}}(q_1) \equiv i \bm{q}_1 \times \bm{\mathsf{V}}(q_1)$ and electric field $\bm{\mathsf{E}}(q_1)=i\omega_1\bm{\mathsf{V}}(q_1)$ on the boundary according to
\begin{align}
  \bm{\mathcal{B}}(q|u) =  \bm{\mathsf{B}}(q)\,\frac{\psi_1(q|u) }{\psi_1(q|0)}  \; ;\qquad \bm{\mathcal{E}}(q|u) = \bm{\mathsf{E}}(q)\,\frac{\psi_1(q|u) }{\psi_1(q|0)}  \;.  \label{eq:30}
\end{align}
Taking the transverse component of (\ref{eq:35}) with respect to $\bm{q}$ and substituting in $f=1-u^2$, we find that
\begin{align}                        
  \bm{\mathtt{V}}_\perp'' - \frac{2u}{1-u^2} \,\bm{\mathtt{V}}_\perp' + \frac{\mathfrak{w}^2 - |\bm{\mathfrak{q}}|^2 (1-u^2)}{u(1-u^2)^2} \bm{\mathtt{V}}_\perp = \frac{\kappa_{_\text{CS}}}{\kappa_{\text{M}}} \bm{G}_\perp(q|u),
  \label{V_perp}
\end{align}
where
\begin{equation}
  \bm{G}_\perp(q|u) = \int \frac{\mathrm{d}^4q_1}{(2\pi)^4} \frac{\mathrm{d}^4q_2}{(2\pi)^4} (2\pi)^4 \delta^4(q_1+q_2-q)\;     \, \bm{\mathcal{G}}_\perp(q_1,q_2|u) 
  \label{G_perp}
\end{equation} 
with
\begin{equation}
  \bm{\mathcal{G}}_\perp^V(q_1,q_2|u)=\bm{\mathcal{G}}^V(q_1,q_2|u)-\frac{\bm{q}}{|\bm{q}|^2}\, \bigg({\bm{q}}\cdot\bm{\mathcal{G}}^V(q_1,q_2|u) \bigg) \;.
  \label{project}
\end{equation}
Eliminating $\bm{\mathtt{V}}_\parallel$ from \eqref{eq:20} and (\ref{eq:22}), we end up with
\begin{equation}
  \mathtt{V}_0''' + \frac{1-3u^2}{u(1-u^2)} \mathtt{V}_0'' + \frac{\mathfrak{w}^2 - |\bm{\mathfrak{q}}|^2 (1-u^2)}{u(1-u^2)^2} \mathtt{V}_0' = \frac{\kappa_{_\text{CS}}}{\kappa_{\text{M}}} M(q|u) \;, \label{eq:37}
\end{equation}
where
\begin{equation}
  M(q|u) \equiv \frac{\mathfrak{w}}{u f^2} G_{_\mathtt{V}}^5(q|u) + \frac{1}{u f} G_{_\mathtt{V}}^0(q|u) = \int \frac{\mathrm{d}^4q_1}{(2\pi)^4} \frac{\mathrm{d}^4q_2}{(2\pi)^4} (2\pi)^4 \delta^4(q_1+q_2-q) \mathcal{M}(q_1,q_2|u) \;,
  \label{M_conv}
\end{equation} 
with
\begin{equation}
\begin{aligned}
  \mathcal{M}(q_1,q_2|u) &\equiv \frac{\mathfrak{w}}{u f^2}\mathcal{G}_V^5(q_1,q_2|u) + \frac{1}{uf} \left( uf \; \mathcal{G}_V^0(q_1,q_2|u) \right)' \\
  &= \frac{3}{\sqrt{2}(\pi T)^2 L} \frac{1}{uf}\; \frac{1}{|\bm{\mathfrak{q}}_2|^2} \bigg( - \mathfrak{w} \, u \, \mathcal{A}_0''(q_2|u)\; \big( \bm{\mathfrak{q}}_2 \cdot \bm{\mathcal{B}}(q_1|u) \big)  \\
  & \hspace{5cm} + \mathfrak{w}_2 \left[ u \mathcal{A}_0'(q_2|u) \; \big(\bm{\mathfrak{q}}_2 \cdot \bm{\mathcal{B}}(q_1|u) \big) \right]' \bigg)  \;.\label{eq:39}
\end{aligned}
\end{equation}
The boundary conditions for $\bm{\mathtt{V}}_\perp$ and $\mathtt{V}_0$ follow from \eqref{eq:36_ren} with $\bm{\mathcal{V}}$ and $\mathcal{V}_0$ replaced with 
$\bm{\mathtt{V}}_\perp$ and $\mathtt{V}_0$, i.e.
\begin{align}
  \bm{\mathtt{V}}_\perp(q|0) = \bm{\mathsf{V}}(q) \;; \qquad  \lim_{u\to 0}  u \mathtt{V}_0''(q|u) = 0 \;. \label{eq:37_ren}
\end{align}
The 2nd equation follows from (\ref{eq:22}) with $\mathsf{V}_0(q)=\bm{q}\cdot\bm{\mathsf{V}}(q)=0$ and the limit  $\lim\limits_{u \to 0}  u \cdot \mathcal{G}_{_\mathrm{V}}^0(q_1,q_2|u) = 0$, the latter of which is evident from the asymptotic behavior of $\bm{\mathcal{V}}$ and $\mathcal{A}_0$ as $u\to 0$.

The solutions of the inhomogeneous equations (\ref{V_perp}) and (\ref{eq:37}) can be constructed from the two pairs of linearly-independent solutions $\{\psi_1, \psi_2\}$ of (\ref{eq:29}) and $\{\phi_1, \phi_1\}$  of (\ref{eq:36}) via the method of variation of parameters with details shown in Appendix~\ref{sec:spec-solut-vari}. The integration constants incurred are fixed by the in-falling wave condition at the horizon and the boundary condition (\ref{eq:36_ren}). We find that
\begin{align}
  \bm{\mathtt{V}}_\perp(q|u) = \bm{C}(q)\, \psi_1(q|u) + \psi_1(q|u) \frac{\kappa_{_\text{CS}}}{\kappa_{\text{M}}} \int_u^1 \ \mathrm{d}\xi\, \frac{\bm{G}_\perp^V(q|\xi)}{W_\perp(\xi)} \psi_2(q|\xi) - \psi_2(q|u) \frac{\kappa_{_\text{CS}}}{\kappa_{\text{M}}} \int_u^1 \ \mathrm{d}\xi\, \frac{\bm{G}_\perp^V(q|\xi)}
  {W_\perp(\xi)} \psi_1(q|\xi) \;, \label{eq:43_perp}
\end{align}
and
\begin{align}
  \mathtt{V}_0'(q|u) = C_0(q)\, \phi_1(q|u) + \phi_1(q|u) \frac{\kappa_{_\text{CS}}}{\kappa_{\text{M}}} \int_u^1 \ \mathrm{d}\xi\, \frac{M(q|\xi)}{W_0(\xi)} \phi_2(q|\xi) - \phi_2(q|u) \frac{\kappa_{_\text{CS}}}{\kappa_{\text{M}}} \int_u^1 \ \mathrm{d}\xi \, \frac{M(q|\xi)}{W_0(\xi)} \phi_1(q|\xi) \;, \label{eq:43_0}
\end{align}
where the constants of integration $\bm{C}(q), C_0(q)$ are given by:
\begin{align}
  \bm{C}(q) &= \frac{\bm{\mathsf{V}}(q)}{\psi_1(q|0)}- \frac{\kappa_{_\text{CS}}}{\kappa_{\text{M}}} \int_0^1 \ \mathrm{d}\xi\, \frac{\bm{G}_\perp^V(q|\xi)}{W_\perp(\xi)} \psi_2(q|\xi)+ \frac{\kappa_{_\text{CS}}}{\kappa_{\text{M}}} \frac{\psi_2(q|0)}{\psi_1(q|0)}\int_0^1 \ \mathrm{d}\xi\, \frac{\bm{G}_\perp^V(q|\xi)}{W_\perp(\xi)} \psi_1(q|\xi) \;, \label{eq:40}\\ 
  C_0(q) &= \frac{\kappa_{_\text{CS}}}{\kappa_{\text{M}}} \int_0^1 \ \mathrm{d}\xi\, \frac{M(q|\xi)}{W_0(\xi)} \phi_2(q|\xi) + \frac{\kappa_{_\text{CS}}}{\kappa_{\text{M}}} \frac{D_2(q)}{D_1(q)}\int_0^1 \ \mathrm{d}\xi\, \frac{M(q|\xi)}{W_0(\xi)} \phi_1(q|\xi) \;, \label{eq:43}
\end{align}
with $D_2(q)=\lim\limits_{u\to 0}u\phi_2'(q|u)$,
and $W_\perp(u)$ and $W_0(u)$ standing for Wronskians of $\{\psi_1,\psi_2\}$ and $\{\phi_1,\phi_2\}$, respectively. It follows from eq.~\eqref{eq:29}\eqref{eq:36} that
\begin{equation}
  \begin{aligned}
    W_\perp(u) &= W[\psi_1, \psi_2] = \frac{\text{const.}}{1-u^2} \;, \\ 
    W_0(u) &= W[\phi_1, \phi_2] = \frac{\text{const.}}{u(1-u^2)} \;,
  \end{aligned}
  \label{wronskian}
\end{equation}
where the overall constants in (\ref{wronskian}) will be cancelled afterwards.

It follows readily from (\ref{eq:43_perp}), (\ref{eq:43_0}) and (\ref{wronskian}) that
\begin{equation}
  \bm{\mathtt{V}}_\perp'(q|0)=\bm{\mathsf{V}}(q)\frac{\psi_1'(q|0)}{\psi_1(q|0)} -\frac{\kappa_{_\text{CS}}}{\kappa_{\text{M}}} \frac{1}{\psi_1(q|0)}\int_0^1 \ \mathrm{d}\xi\, (1-\xi^2) \bm{G}_\perp^V(q|\xi) \psi_1(q|\xi)  \label{J_perp}
\end{equation}
and
\begin{equation}
  \mathtt{V}_0'(q|0) = \frac{\kappa_{_\text{CS}}}{\kappa_{\text{M}}} \frac{1}{ D_1(q)}\int_0^1 \ \mathrm{d}\xi\, \xi(1-\xi^2)M(q|\xi)\phi_1(q|\xi) \label{J_0} \;.
\end{equation}
Substituting (\ref{J_perp}) and (\ref{J_0}) into the continuity equation (\ref{eq:20}), the derivative of the longitudinal component of $\bm{\mathtt{V}}(q|0)$ with respect to $u$, $\bm{\mathtt{V}}_\parallel'(q|0)$, is obtained and we have
\begin{equation}
  \bm{\mathtt{V}}'(q|0) = \bm{\mathtt{V}}_\perp'(q|0) - \frac{\mathfrak{w}\, \bm{\mathfrak{q}}}{|\bm{\mathfrak{q}}|^2} \mathtt{V}_0'(q|0)  - \frac{\kappa_{_\text{CS}}}{\kappa_{_\text{M}}} \frac{ \bm{\mathfrak{q}}}{|\bm{\mathfrak{q}}|^2} G^5_{\mathtt{V}}(q|0) \;.
  \label{total_V}
\end{equation}
It follows from (\ref{eq:8}) and  $\mathtt{F}_{\mu 5}= \mathtt{V}_\mu'$ that the vector current of the boundary field theory reads
\begin{equation}
  \bm{J}(q)=-2\pi^2T^2\bm{\mathtt{V}}'(q|0)-3\sqrt{2}\kappa_{_\text{CS}}\int \frac{\mathrm{d}^4q_1}{(2\pi)^4} \frac{\mathrm{d}^4 q_2}{(2\pi)^4} (2\pi)^4 \delta^4(q_1+q_2-q) \, \mathtt{A}_0(q_2)\bm{\mathsf{B}}(q_1) \;.
  \label{total_current}
\end{equation}
The $\mathcal{O}(1)$  term of $\bm{\mathtt{V}}'(q|0)$ above, i.e., the first term of (\ref{J_perp}), contributes to the polarization current calculated in \cite{Policastro2002a} and the $\mathcal{O}(\kappa_{_\text{CS}})$ terms give rise to the chiral magnetic current that is the theme of the next section.

% %%%%%% 
\section{Chiral Magnetic Current}
\label{sec:smallq}

With the formulation developed in the preceding section, we are equipped to calculate the chiral magnetic current for arbitrarily spacetime-dependent chiral imbalance and magnetic field in this section. The chiral imbalance consists of a net axial charge characterized by a constant axial chemical potential $\mu_A$ and its spacetime variation proxied by the temporal component of a spacetime dependent axial vector potential $\mathsf{A}_0$. Adapting the $U_A(1)$ gauge invariant definition of $\mu_A$ in the holographic environment, we impose the condition that the Fourier component of $\mathsf{A}_0$ with zero spatial momentum vanishes on the AdS-boundary, e.g. \eqref{eq:55}, which implies that  
\begin{equation}
\frac{1}{\Omega}\int \mathrm{d}\bm{r}\, \mathsf{A}_0(x)=0 \;,
\end{equation}
with $\Omega$ the spatial volume of the system. Correspondingly, the CME current consists of the contribution from the axial chemical potential and that from the three-point function, i.e.
\begin{equation}
  \bm{J}_{\rm CME}(q) = \bm{\mathsf{J}}(q) + \bm{J}_{\rm AVV}(q) \;,
  \label{cme_current}
\end{equation}
with both terms proportional to the anomaly coefficient $\mathcal{C}$. 

To simplify the notations, we suppress the subscript of the retarded solutions $\phi_1$ and $D_1(q)$, i.e.
\begin{equation}
  \phi(q|u) \equiv \phi_1(q|u) \;;\qquad  D(q) \equiv D_1(q) = \lim_{u\to 0}\, u\phi_1'(q|u)
  \label{eq:56}
\end{equation}
and introduce
\begin{equation}
  \psi(q|u) \equiv \frac{\psi_1(q|u)}{\psi_1(q|0)}  \;.
\end{equation}

The first term of (\ref{cme_current}) follows readily from (\ref{G_mu_A}), we have
\begin{equation}
  \bm{\mathsf{J}}(q) =3\sqrt{2}\kappa_{_\text{CS}} \mu_A\int_0^1 \mathrm{d}u \, \bm{\mathcal{B}}(q|u)\psi(q|u) = \mu_A K(q)\bm{\mathsf{B}}(q)
  \label{CME_general}
\end{equation}
with the kernel
\begin{equation}
  K(q) = 3\sqrt{2}\kappa_{_\text{CS}} \int_0^1 \, \mathrm{d}u \; \psi^2(q|u) \;.
  \label{kernelK}
\end{equation}
The AVV contribution can be obtained by substituting  (\ref{eq:25}), (\ref{G_perp}) and (\ref{M_conv}) into (\ref{J_perp}), (\ref{J_0}) and (\ref{total_V}) together with explicit expressions (\ref{eq:31}) , (\ref{eq:35}) , (\ref{project}) and (\ref{eq:39}), and then collecting the $O(\kappa_{_\text{CS}})$ terms of (\ref{total_current}). Finally, we find that
\begin{equation}
  J_{\rm AVV}^\mu(q) = \int \frac{\mathrm{d}^4q_1}{(2\pi)^4} \frac{\mathrm{d}^4q_2}{(2\pi)^4} (2\pi)^4 \delta^4(q_1+q_2-q)\mathcal{J}^\mu(q_1,q_2) \;, \label{eq:58}
\end{equation}
with $q_1$ and $q_2$ the 4-momenta carried by the magnetic field and the axial-vector potential, respectively. As a result, the spatial component reads
\begin{equation}
\begin{aligned}
  \bm{\mathcal{J}}(q_1,q_2) &= 2 (\pi T)^2 L \kappa_{_\text{CS}} \bigg[ \int_0^1 \ \mathrm{d}u\, (1-u^2) \bm{\mathcal{G}}_\perp^V(q_1,q_2|u) \; \psi(q|u) \\
   & \hspace{3cm} + \frac{\mathfrak{w}\, \bm{\mathfrak{q}} }{|\bm{\mathfrak{q}} |^2D(q)} \int_0^1 \ \mathrm{d}u\, u(1-u^2)\mathcal{M}(q_1,q_2|u) \; \phi(q|u) \bigg]  \\
  & \hspace{4cm}  + 3\sqrt{2} \kappa_{_\text{CS}}\, \frac{\bm{\mathfrak{q}}}{|\bm{\mathfrak{q}}|^2}\mathtt{A}_0(q_2) \big( \bm{\mathfrak{q}}_2\cdot\bm{\mathsf{B}}(q_1) \big) - 3\sqrt{2} \kappa_{_\text{CS}}\,\mathtt{A}_0(q_2) \bm{\mathsf{B}}(q_1) \;,
\end{aligned}
\label{avv_current}
\end{equation}
where the magnetic Gauss law $\bm{\mathfrak{q}}_1\cdot\bm{\mathsf{B}}(q_1) =0$ is employed so $\bm{\mathfrak{q}}\cdot\bm{\mathsf{B}}(q_1) =\bm{\mathfrak{q}}_2\cdot\bm{\mathsf{B}}(q_1)$. The temporal component of the AVV current is
\begin{align}
  \mathcal{J}^0(q_1,q_2) &= \frac{2 (\pi T)^2L \kappa_{_\text{CS}}}{D(q)} \int_0^1 \ \mathrm{d} u\, u(1-u^2)\mathcal{M}(q_1,q_2|u)\phi_1(q|u)  \;.
                           \label{avv_charge}
\end{align}
which represents the charge induced by a spacetime-dependent chiral imbalance.
As shown in Appendix.~\ref{sec:heun}, $\psi_1(q|u)$ reduces to a hypergeometric function for a homogeneous magnetic field, i.e. $q=(\omega,0)$ and (\ref{CME_general}) becomes
\begin{equation}
  \small{
    \bm{\mathsf{J}}(q) = 3\sqrt{2}\kappa_{_\text{CS}}  \mu_A \, \bm{\mathsf{B}}(q)  \frac{\Gamma^2 (\frac{1-i\mathfrak{w}}{2} )\Gamma^2 (\frac{3-i\mathfrak{w}}{2})}{\Gamma^2 (1-i\mathfrak{w} )} \int_0^1 \mathrm{d}u\, \bigg[ \left(\frac{1-u}{1+u} \right)^{-i\frac{\mathfrak{w}}{2}}F \left(\frac{1-i}{2}\mathfrak{w},-\frac{1+i}{2}\mathfrak{w};1-i\mathfrak{w};\frac{1-u}{1+u} \right) \bigg]^2 \;. \label{eq:59}
    }
\end{equation}
Phenomenologically, a homogeneous magnetic field serves a reasonable approximation for a sufficiently small fireball in RHIC. While the $\mu_A$ contribution to the chiral magnetic current (\ref{CME_general}) supports a power series expansion in momentum $q$ (equivalently, gradient expansion in coordinate space), the low momenta (long wavelength) behavior of the AVV contribution (\ref{avv_current}) is far from trivial because of the diffusion denominator $D(q_2)$ brought about by $\mathsf{A}_0(q_2)$, which is hidden in the integrand of (\ref{avv_current}). The same $\mathsf{A}_0(q_2)$ is also responsible to the diffusion denominator $D(q)$ pertaining to the longitudinal component of the AVV current (\ref{avv_current}) through the induced charge density (\ref{avv_charge}).

The low momentum expansion of the solutions $\psi(q|u)$ and  $\phi(q|u)$ can be obtained by the transformation
\begin{equation}
  \begin{aligned}
    \psi(q|u) &=(1-u)^{-i\frac{\mathfrak{w}}{2}} G(q|u)    \;, \\ 
    \phi(q|u) &=(1-u)^{-i\frac{\mathfrak{w}}{2}} F(q|u)  \;,
  \end{aligned}
  \label{eq:48}
\end{equation}
and the equations (\ref{eq:29}) and (\ref{eq:36}) become
\begin{align}
  G''+ \left(-\frac{2u}{1-u^2}+i\frac{\mathfrak{w}}{1-u}\right)G'+\left[\frac{i\mathfrak{w}}{2(1-u^2)}+
  \frac{\mathfrak{w}^2(4+3u+u^2)}{4u(1+u)(1-u^2)}-\frac{ |\bm{\mathfrak{q}}|^2}{4u(1-u^2)}\right]G &=0  \;, \label{eq:45} \\ 
  F'' + \left[\frac{1-3u^2}{u(1-u^2)} + \frac{ i \mathfrak{w}}{1-u}\right]F'+
  \left[i\mathfrak{w} \, \frac{(1+2u)}{2u(1-u^2)} + \mathfrak{w}^2 \, \frac{(4+3u+u^2)}{4u(1+u)(1-u^2)}-\frac{|\bm{\mathfrak{q}}|^2}{u(1-u^2)}\right]F &=0 \;. \label{eq:46} 
\end{align}
Moving the terms containing $\mathfrak{w}$ and $\bm{\mathfrak{q}}$ on LHS of eqs.~\eqref{eq:45}\eqref{eq:46} to RHS and solving the equations iteratively starting with the leading order solutions 
$F^{(0)}=1$ and $G^{(0)}=1$, we derive, to the order we need in this section, that
\begin{align}
  G(q|u) &= 1+i\frac{\mathfrak{w}}{2}\ln\frac{1+u}{2}+\frac{ |\bm{\mathfrak{q}}|^2}{2}\left[\frac{\pi^2}{12}+{\rm{Li}}_2(-u)+\ln u\ln(1+u)+{\rm{Li}}_2(1-u)\right] + \cdots \;;  \label{eq:44}\\ 
  F(q|u) &= 1+i\frac{\mathfrak{w}}{2}\ln\frac{2u^2}{1+u} + |\bm{\mathfrak{q}}|^2\ln\frac{1+u}{2u}+ \cdots \;, \label{eq:47}
  % \label{leadingsol}
\end{align}
with $\rm{Li}_2(u)$ the Spence function. Combining with the expansion: $(1-u)^{-i\frac{\mathfrak{w}}{2}}=1-i\frac{\mathfrak{w}}{2}\ln(1-u) + \cdots $, we end up with
\begin{align}
  \psi(q|u) &= 1+i\frac{\mathfrak{w}}{2}\ln\frac{1+u}{1-u} +  \frac{|\bm{\mathfrak{q}}|^2}{2}\left[ - \frac{\pi^2}{6} + {\rm{Li}}_2(-u) + {\rm{Li}}_2(1-u) + \ln u \ln(1+u) \right] + \cdots   \;,  \label{eq:50}\\ 
  \phi(q|u) &= 1 + i\frac{\mathfrak{w}}{2}\ln\frac{2u^2}{1-u^2} + |\bm{\mathfrak{q}}|^2 \ln\frac{1+u}{2u}+ \cdots \;,  \label{eq:51}
\end{align}
It follows from the definition \eqref{eq:38} that
\begin{equation}
  D(q)=  \underbrace{ i\, \mathfrak{w}- |\bm{\mathfrak{q}}|^2}_{D_{(0)}(q)} + \underbrace{  \left[ \frac{\mathfrak{w}^2}{2} + \frac{i}{2}\mathfrak{w}|\bm{\mathfrak{q}}|^2 - (|\bm{\mathfrak{q}}|^2)^2 \right]\ln 2 }_{D_{(1)}(q)} + \cdots \;.
  \label{Dq}
\end{equation} 
Eqs.~\eqref{eq:44}\eqref{eq:47} and the first two terms of the (\ref{Dq}), $D_{(0)}(q)$, were derived in Ref.\cite{Policastro2002a} in the context of two-point functions. We derive the higher order terms, $D_{(1)}(q)$, of eq.(\ref{Dq}) in Appendix \ref{sec:diffusion}.

In what follows, we shall apply the low momenta expansions of (\ref{eq:50}) and (\ref{eq:51}) for $\psi(q|u)$ and $\phi(q|u)$ to the AVV current in the form $\mathcal{J}^\mu(q_1,q_2) =  \Lambda^{\mu\nu\rho}(q_1,q_2)\mathsf{A}_\rho(q_2)\mathsf{V}_\nu(q_1)$. To highlight the role of the diffusion denominator \eqref{Dq}, the orders of these expansions are sorted by scaling the dimensionless momenta of the kernel  $\Lambda^{\mu\nu\rho}(q_1,q_2)$ according to 
\begin{equation}
  \mathfrak{w}_{1,2} \to \lambda\, \mathfrak{w}_{1,2}\qquad \bm{\mathfrak{q}}_{1,2} \to \sqrt{\lambda}\, \bm{\mathfrak{q}}_{1,2} \;. \label{eq:41}
\end{equation}
So $D_{(0)}(q)$ in eq.~\eqref{Dq} contributes to the leading power of $\lambda$, while $D_{(1)}(q)$ to the subleading power. Consequently, the three-point function $\Lambda^{\mu\nu\rho}(q_1,q_2)$ will be expanded in powers of $\lambda$, its leading order and the subleading order of momenta described below correspond to the leading power and subleading power in $\lambda$, respectively. 
The scale factor $\lambda$ is set to one in the end for low momenta, hence $\lambda$ will be omitted in all expressions below.

\subsection{Leading Order}
\label{sec:leading-order}

The leading order contribution to the current is given by the $O(1)$ terms of $\psi(q|u)$, $\phi(q|u)$ in \eqref{eq:50}\eqref{eq:51} and $O(\mathfrak{w}, \bm{\mathfrak{q}}^2)$ term of $D(q)$ in \eqref{Dq}, i. e. 
\begin{align}
  \psi(q|u) \simeq 1, \qquad \phi(q|u) \simeq 1 \;,
  \label{leadingsol}
\end{align} 
and $D(q)\simeq D_{(0)}(q) = i\mathfrak{w}-|\bm{\mathfrak{q}}|^2$ throughout this subsection. Substituting these approximations to \eqref{free}, we have
\begin{align}
  \bm{\mathcal{V}}_\perp(q|u) \simeq \bm{\mathsf{V}}(q) \;,\qquad \mathcal{V}_0'(q|u)=0 \;, \qquad 
  \mathcal{A}_0'(q|u) \simeq |\bm{\mathfrak{q}}|^2\frac{ \mathsf{A}_0(q)}{D_{(0)}(q)} \;.
  \label{zerothsol} 
\end{align}
It follows from eq.(\ref{zerothsol}), (\ref{eq:31}), (\ref{eq:33}) and (\ref{eq:35}) that to the leading order
\begin{align}
  \bm{\mathcal{G}}_V(q_1,q_2|u) &\simeq - \frac{1}{(1-u^2)}\, \frac{3\, |\bm{\mathfrak{q}}_2|^2 \mathsf{A}_0(q_2)}{\sqrt{2}(\pi  T)^2 L  \, D_{(0)}(q_2)} \bm{\mathsf{B}}(q_1)  \;, \label{G_V_0} \\
  \bm{\mathcal{G}}_\perp(q_1,q_2|u) &\simeq - \frac{1}{(1-u^2)}\,\frac{3\, |\bm{\mathfrak{q}}_2|^2 \mathsf{A}_0(q_2)}{\sqrt{2} (\pi  T)^2 L \, D_{(0)}(q_2)} \left[\bm{\mathsf{B}}(q_1)-\frac{\bm{\mathfrak{q}}}{ |\bm{\mathfrak{q}}|^2} \big( \bm{\mathfrak{q}}_2\cdot\bm{\mathsf{B}}(q_1) \big) \right]   \;,\label{G_V_0_tr} \\
  \mathcal{M}(q_1,q_2|u) &\simeq \frac{1}{u(1-u^2)} \frac{3\mathfrak{w}_2\mathsf{A}_0(q_2)}{\sqrt{2} (\pi T)^2 L \, D_{(0)}(q_2) } \big( \bm{\mathfrak{q}}_2\cdot\bm{\mathsf{B}}(q_1) \big) \;.  \label{M_0} 
\end{align}
Substituting (\ref{G_V_0_tr}) and (\ref{M_0}) 
into (\ref{avv_current}), we obtain the leading order CME current in terms of the axial gauge potential and magnetic field:
\begin{equation}
  \bm{\mathcal{J}}_{(0)}(q_1,q_2) = - 3\sqrt{2}\kappa_{_\text{CS}} \frac{ \mathsf{A}_0(q_2)}{D_{(0)}(q_2)}\, i \mathfrak{w}_2\, \bigg[ \bm{\mathsf{B}}(q_1) + \frac{\bm{\mathfrak{q}}}{D_{(0)}(q)} \big(\bm{\mathfrak{q}}_2\cdot \bm{\mathsf{B}}(q_1) \big) \bigg] \label{leading-current} .
\end{equation}
The corresponding charge density follows from (\ref{M_0}) and (\ref{avv_charge}), and reads
\begin{equation}
  \mathcal{J}^0_{(0)}(q_1,q_2) = 3\sqrt{2}\kappa_{_\text{CS}}\, \frac{\mathsf{A}_0(q_2)\, \big(\bm{\mathfrak{q}}_2\cdot\bm{\mathsf{B}}(q_1) \big)}{D_{(0)}(q)D_{(0)}(q_2)}  \mathfrak{w}_2
  \;.
  \label{leading-charge}
\end{equation}

The non-local response because of the diffusion denominators $D_{(0)}(q)$ and $D_{(0)}(q_2)$ in the eq.~\eqref{leading-current} underlies nontrivial infrared behaviors of  of $\bm{J}_{\rm AVV}(q)$ through eq.~\eqref{eq:58}. For the case $|\bm{\mathfrak{q}}_2|^2 \ll \mathfrak{w}_2 \ll 1$, we have
\begin{equation}
  \bm{\mathcal{J}}_{(0)}(q_1,q_2) \simeq - 3\sqrt{2}\kappa_{_\text{CS}} \mathsf{A}_0(q_2)\bm{\mathsf{B}}(q_1) \;,
  \label{IR_1}
\end{equation}
while in the opposite case $\mathfrak{w}_2 \ll |\bm{\mathfrak{q}_2}|^2 \ll 1$ ,
\begin{equation}
  \bm{\mathcal{J}}_{(0)}(q_1,q_2) \simeq 0   \;.
  \label{IR_2}
\end{equation}
From the convolution eq.~\eqref{eq:58}, the two quantities $\bm{\mathfrak{q}}\times \bm{\mathcal{J}}_{(0)}(q_1,q_2)$ and $\bm{\mathfrak{q}} \cdot \bm{\mathcal{J}}_{(0)}(q_1,q_2)$ respond to the transverse and longitudinal components of CME spatial current $\bm{J}_{_\mathrm{AVV}}(q)$, respectively. We investigate the non-trivial infrared behavior of the current $\bm{J}_{_\mathrm{AVV}}(q)$ by numerically plotting the relations between the these components of $\bm{J}_{_\mathrm{AVV}}(q)$ and the small $q$'s.

\begin{figure}[H]
  \centering{\includegraphics[width=13cm]{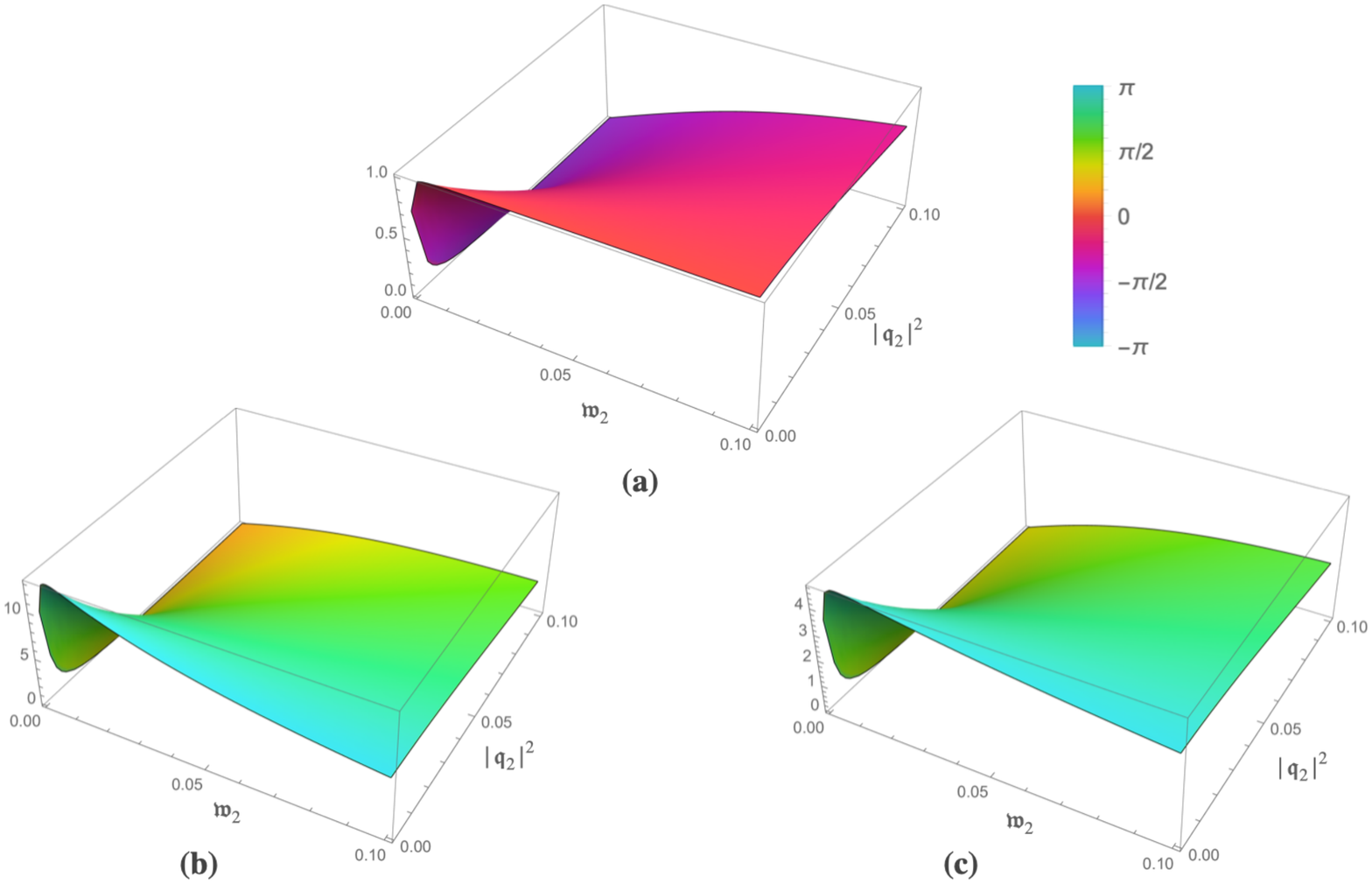}}
  \caption{As an illustration of the non-local response of the vector current, we project out its transverse and longitudinal parts as $\bm{\mathfrak{q}}\times\bm{\mathcal{J}}_{(0)}(q_1,q_2) \equiv C_\perp(q_1,q_2)\, \mathsf{A}_0(q_2)\,  \bm{\mathfrak{q}}\times\bm{\mathsf{B}}(q_1)$ and $\bm{\mathfrak{q}} \cdot \bm{\mathcal{J}}_{(0)}(q_1,q_2)\equiv C_\parallel(q_1,q_2)\, \mathsf{A}_0(q_2) \bm{\mathfrak{q}} \cdot \mathsf{B}(q_1)$, and plot the coefficients $C_\perp(q_1,q_2)$ in fig.(a) and $C_\parallel(q_1,q_2)$ in fig.(b) and fig.(c) to the leading order in small momenta. Since the longitudinal component involves two diffusion denominators $D_{(0)}(q_1)$ and $D_{(0)}(q_2)$, however, the momenta $q_1$ doesn't influence the infrared behaviour for $D_{(0)}(q_2)$, as is demonstrated by fig.(b) with $\mathfrak{w}_1=0.07; |\bm{\mathfrak{q}}_1|=0.02$, and fig.(c) with  $\mathfrak{w}_1=0.2; |\bm{\mathfrak{q}}_1|=0.12$ (the angle between $\bm{\mathfrak{q}}_1$ and $\bm{\mathfrak{q}}_2$ takes $\frac{\pi}{3}$). The expression can be read off from eq.~\eqref{leading-current}. The subtlety of the infrared limits reflected in eqs.\eqref{IR_1} and \eqref{IR_2} is manifested in the plot.}
  \label{fig:leading-term}
\end{figure}

The charts in Fig.~\ref{fig:leading-term} display visually the leading order contribution from the AVV three-point function on CME current, and in the limit $(\mathfrak{w}_2, \bm{\mathfrak{q}}_2 ) \to 0$, the two components $\bm{J}_{_\mathrm{AVV}}(q)_\parallel$ and $\bm{J}_{_\mathrm{AVV}}(q)_\perp$ behave in the manner reflected by the analytical results eq.~\eqref{IR_1} and eq.~\eqref{IR_2}.

As shown at the end Section \ref{sec:action}, the asymptotic behavior (\ref{IR_1}) is a direct consequence of the anomalous Ward identity \eqref{eq:11}, and its validity is not limited to small $\mathfrak{w}_2$. The asymptotic behavior (\ref{IR_2}) is a holographic version of the Coleman-Hill theorem and implies null chiral magnetic current at $\mu_A=0$, in agreement with the conclusion of \cite{Rubakov2010} for a simplified holographic model. Both asymptotic behaviors match the field theoretic result of the AVV three-point function\cite{Hou}. It follows from (\ref{cme_current}), (\ref{CME_general}), (\ref{eq:58}) and (\ref{IR_1}) that for a $\mathsf{A}_0(x)$ that varies slowly in space, the AVV three point function can significantly contribute to the chiral magnetic current with $\mathsf{A}_0(x)\sim \mu_A$.

Restoring the dimensions of all 4-momenta via (\ref{eq:dimensionless}) and substituting in $\mathsf{B}_i(q_1)=i\epsilon_{ikj}q_{1k}\mathsf{V}_j(q_1)$ together with the relation between the Chern-Simons coupling and anomaly coefficient (\ref{anomaly_coeff}), we extract the leading order AVV function exhibited in (\ref{J_spatial}) and (\ref{charge_0}).

% %%%%%%%%%%%%%%
\subsection{Subleading Order}
\label{sec:subleading-order}

In order to obtain the subleading order of the vector current $\mathcal{J}^\mu_{(1)} = \{ \mathcal{J}^0_{(1)}(q_1, q_2|u) , \bm{\mathcal{J}}_{(1)}(q_1, q_2|u) \}$, we need to include the  $O(\mathfrak{w}, |\bm{\mathfrak{q}}|^2)$ terms in $\phi(q|u)$, $\psi(q|u)$ and the $O(\mathfrak{w}^2, \mathfrak{w}\,|\bm{\mathfrak{q}}|^2, (|\bm{\mathfrak{q}}|^2)^2)$ terms in $D(q)$ shown in \eqref{eq:50}\eqref{eq:51} and \eqref{Dq}, respectively. Substituting \eqref{eq:50}\eqref{eq:51} into \eqref{free} and taking the derivatives with respect to $u$, we find that
\begin{align}
  \mathcal{A}_0'(q|u) &= \frac{\mathsf{A}_0(q)}{D(q)} |\bm{\mathfrak{q}}|^2 \bigg[1 + \frac{i \mathfrak{w}}{2} \ln \frac{2u^2}{1-u^2} + |\bm{\mathfrak{q}}|^2 \ln \frac{1+u}{2u}  \cdots\bigg]  \label{eq:18}\\
  \mathcal{A}_0''(q|u) &= \frac{\mathsf{A}_0(q)}{D(q)} |\bm{\mathfrak{q}}|^2 \bigg[ \frac{i \mathfrak{w}}{u(1-u^2)} - \frac{|\bm{\mathfrak{q}}|^2}{u(1+u)} + \cdots \bigg]   \label{eq:26}\\
  \bm{\mathcal{V}}(q|u) & = \bm{\mathsf{V}}(q) \, \bigg[ 1 + \frac{i \mathfrak{w}}{2} \ln \frac{1+u}{1-u} + \frac{|\bm{\mathfrak{q}}|^2}{2} \left( -\frac{\pi^2}{6} + \mathrm{Li_2}(-u) + \mathrm{Li_2}(1-u) + \ln u \ln (1+u) \right)  \cdots \bigg]   \label{eq:27}\\
  \bm{\mathcal{V}}'(q|u) & = \frac{\bm{\mathsf{V}}(q)}{1-u^2} \, \bigg[  i\mathfrak{w}+ |\bm{\mathfrak{q}}|^2\ \ln u + \cdots  \bigg] \;, \label{eq:28}
\end{align}
with $D(q) \simeq D_{(0)}(q) + D_{(1)}(q)$, as shown in \eqref{Dq}.

Substituting (\ref{eq:18}), (\ref{eq:26}), (\ref{eq:27}) and (\ref{eq:28}) into (\ref{eq:35}), (\ref{G_perp}) and (\ref{eq:39}), and then carrying out the integrations in (\ref{avv_current}), we obtain, to the subleading order, that
\begin{align}
  \int_0^1 \mathrm{d}u \; (1-u^2) \bm{\mathcal{G}}_\perp(q_1,q_2|u) \; \psi(q|u) &= -\frac{3}{\sqrt{2} (\pi T)^2 L} \frac{\mathsf{A}_0(q_2)}{D(q_2)} \Bigg\{ |\bm{\mathfrak{q}}_2|^2 \; \bm{\mathsf{B}}(q_1)\, + \bm{\mathsf{S}}(q_1,q_2)   \notag \\ 
                                                                                 & \qquad - \frac{\bm{\mathfrak{q}}}{|\bm{\mathfrak{q}}|^2}\bigg[ \bm{\mathfrak{q}} \cdot\left(  |\bm{\mathfrak{q}}_2|^2 \; \bm{\mathsf{B}}(q_1)\, + \bm{\mathsf{S}}(q_1,q_2) \right) \bigg]  \cdots \Bigg\}  \label{eq:42}\\
  \int_0^1 \mathrm{d}u \; u(1-u^2)\, \mathcal{M}(q_1,q_2|u) \phi(q|u)  &= \frac{3}{\sqrt{2}}\frac{1}{(\pi T)^2 L} \frac{\mathsf{A}_0(q_2)}{D(q_2)} \big(\bm{\mathfrak{q}}\cdot \bm{\mathsf{B}}(q) \big) \bigg[\mathfrak{w}_2  + \mathsf{S}(q_1,q_2) \cdots \bigg]   \label{eq:49} \;,
\end{align}
where
\begin{align}
  \bm{\mathsf{S}}(q_1,q_2) &=  \bigg[ \left(2 i \mathfrak{w}_1 + \frac{i}{2} \mathfrak{w}_2 + |\bm{\mathfrak{q}}_2|^2 \right) \ln 2 -\frac{\pi^2}{12} \left(|\bm{\mathfrak{q}_1}|^2 + |\bm{\mathfrak{q}}|^2 \right)  \bigg] |\bm{\mathfrak{q}}_2|^2\, \bm{\mathsf{B}}(q_1)  \notag \\
                           & \hspace{1cm} + \left( - \frac{\pi^2}{8} i \mathfrak{w}_2 + \frac{\pi^2}{12} |\bm{\mathfrak{q}}_2|^2 \right)\; \bm{\mathfrak{q}}_2 \times \big( \bm{\mathfrak{q}}_1 \times \bm{\mathsf{B}}(q_1) \big) + i \ln 2 \, |\bm{\mathfrak{q}}_2|^2 \left( \bm{\mathfrak{q}_2} \times \bm{\mathsf{E}}(q_1) \right)  \;, \label{eq:subleading-1}\\
  \mathsf{S}(q_1,q_2) &= -\frac{i}{2} \mathfrak{w}_2^2 \ \ln 2 - \frac{i}{2}  \, \mathfrak{w}_2\mathfrak{w} \ \ln 2 + (\mathfrak{w} |\bm{\mathfrak{q}}_2|^2 + \mathfrak{w}_2 |\bm{\mathfrak{q}}|^2 ) \ln 2 - \frac{\pi^2}{8} \mathfrak{w}_2 |\bm{\mathfrak{q}}_1|^2 \;. \label{eq:subleading-2} 
\end{align}
It follows from (\ref{avv_current}), (\ref{avv_charge}), (\ref{eq:49}), (\ref{eq:subleading-1}) and (\ref{eq:subleading-2}) that the AVV contribution to the CME current up to the subleading order reads
\begin{equation}
  \begin{aligned}
    \bm{\mathcal{J}}(q_1,q_2) &\simeq \bm{\mathcal{J}}_{(0)}(q_1,q_2) + \bm{\mathcal{J}}_{(1)}(q_1,q_2)  \\
    &= - 3\sqrt{2}\kappa_{_\text{CS}} \frac{\mathsf{A}_0(q_2)}{D_{(0)}(q_2)}\, \Bigg\{ i \mathfrak{w}_2\, \bigg[ \bm{\mathsf{B}}(q_1) + \frac{\bm{\mathfrak{q}}}{D_{(0)}(q)} \Big(\bm{\mathfrak{q}}_2\cdot \bm{\mathsf{B}}(q_1) \Big) \bigg]  \\
    &\quad  - \bigg[ \bigg(  2\, \mathfrak{w}_1 \mathfrak{w}_2 +  \mathfrak{w}_2^2 + 2i \, \mathfrak{w}_1 |\bm{\mathfrak{q}}_2|^2 \bigg)\frac{\ln 2}{D_{(0)}(q_2)}  + \frac{\pi^2}{12}\left(|\bm{\mathfrak{q}_1}|^2 + |\bm{\mathfrak{q}}|^2 \right)  \bigg]\,  |\bm{\mathfrak{q}}_2|^2 \, \bm{\mathsf{B}}(q_1)  \\
    &\quad  - \Bigg[ \frac{\pi^2}{12}|\bm{\mathfrak{q}_2}|^2  + \left( \frac{\mathfrak{w}^2}{2} + \frac{i \mathfrak{w}}{2}|\bm{\mathfrak{q}}|^2 - (|\bm{\mathfrak{q}}|^2)^2 \right) \frac{\ln 2}{D_{(0)}(q)} + \left(\frac{\mathfrak{w}_2^2}{2} + \frac{i \mathfrak{w}_2}{2}|\bm{\mathfrak{q}}_2|^2 - (|\bm{\mathfrak{q}}_2|^2)^2 \right) \frac{\ln 2}{D_{(0)}(q_2)}  \\ 
    & \quad\qquad  + \left(\frac{1}{2}\mathfrak{w}_1\mathfrak{w}_2\ln 2 + i\mathfrak{w}_2|\bm{\mathfrak{q}}_2|^2\ln 2 - \mathfrak{w}|\bm{\mathfrak{q}}|^2\ln 2 + \frac{\pi^2}{8}\, i \mathfrak{w}_1|\bm{\mathfrak{q}}_1|^2\right) \Bigg] \frac{\bm{\mathfrak{q}}}{D_{(0)}(q)} \big(\bm{\mathfrak{q}}_2\cdot \bm{\mathsf{B}}(q_1) \big) \\
    &\quad  - \left( - \frac{\pi^2}{8} i \mathfrak{w}_2 + \frac{\pi^2}{12} |\bm{\mathfrak{q}}_2|^2 \right)\; \bm{\mathfrak{q}}_2 \times \Big( \bm{\mathfrak{q}}_1 \times \bm{\mathsf{B}}(q_1) \Big) - i \ln 2 \, |\bm{\mathfrak{q}}_2|^2 \Big( \bm{\mathfrak{q}_2} \times \bm{\mathsf{E}}(q_1) \Big) \ \Bigg\} \;.
  \end{aligned}\label{eq:53}
\end{equation}
The first line above is the leading order of the CME current \eqref{leading-current}, $\bm{\mathcal{J}}_{(0)}(q_1,q_2)$, and the diffusion denominators $D_{(0)}(q)$ and $D_{(0)}(q_2)$ are the first term of (\ref{Dq}) so that the entire expression is accurate to the required order. The presence of $|\bm{\mathfrak{q}}|^2$ in the denominators of the formulas~\eqref{avv_current} and~\eqref{project} gives rise to the direction singularity characterised by $(\mathfrak{q}_i\mathfrak{q}_j)/|\bm{\mathfrak{q}}|^2$. However, like the leading order result~\eqref{leading-current}, the $1/|\bm{\mathfrak{q}}|^2$-term is eventually cancelled in the subleading order. Though it is not obvious yet, we suspect that this cancellation is generic, not limited to the small momenta. Extracting the AVV function from \eqref{eq:53} and take the limit $\bm{\mathfrak{q}}_2\to 0$, we found that the same infrared limit, Eqn.~\eqref{order_1}, emerges as expected.

   The corresponding charge density follows readily from (\ref{avv_charge}), (\ref{eq:49}) and (\ref{eq:subleading-2}) and we have explicitly that
\begin{align}
  \mathcal{J}^0(q_1,q_2) &= 3 \sqrt{2} \kappa_{_\text{CS}} \, \frac{\mathsf{A}_0(q_2)\, \big( \bm{\mathfrak{q}} \cdot \bm{\mathsf{B}}(q_1) \big)}{D_{(0)}(q_2) D_{(0)}(q)} \Bigg[ \mathfrak{w}_2  -  \left( \frac{\mathfrak{w}^2}{2} + \frac{i}{2}\mathfrak{w}|\bm{\mathfrak{q}}|^2 - (|\bm{\mathfrak{q}}|^2)^2 \right) \frac{\ln 2}{D_{(0)}(q)} \notag \\
  &\qquad \qquad 
   - \left(\frac{\mathfrak{w}_2^2}{2} + \frac{i}{2}\mathfrak{w}_2|\bm{\mathfrak{q}}_2|^2 - (|\bm{\mathfrak{q}}_2|^2)^2 \right) \frac{\ln 2}{D_{(0)}(q_2)}  \notag\\
 & \qquad \qquad  -\frac{i}{2} \mathfrak{w}_2^2 \ \ln 2 - \frac{i}{2}  \, \mathfrak{w}_2\mathfrak{w} \ \ln 2 +  \big(\mathfrak{w} |\bm{\mathfrak{q}}_2|^2 + \mathfrak{w}_2 |\bm{\mathfrak{q}}|^2 \big) \ln 2 - \frac{\pi^2}{8} \mathfrak{w}_2 |\bm{\mathfrak{q}}_1|^2  \Bigg] \;.  \label{eq:54}
\end{align}

Before concluding this section, we would like to comment on the relativistic causality. As discussed above, the $1/|\mathfrak{q}|^2$ factor is expected to be cancelled so there is no action at a distance. While the diffusion denominator $D(q)$ appears supporting superluminal response to the external sources, this is an artifact of the low momenta expansion. Low momenta $(\omega, \bm{q})$ correspond to the large spacetime separation $(\Delta t, \Delta\bm{r})$ between a signal origination and its detection where the diffusion profile $|\Delta\bm{r}|\sim\sqrt{\Delta t}$ for $\Delta t>0$ is well within the future light cone $|\Delta\bm{r}|=|\Delta t|$, and thus, the response is causal. The bottom line is that the Maxwell-Chern-Simons equation we are solving is a set of classical field equations in a curved background without curvature singularity (the outside of the horizon). So, by means of the equivalence principle, there exist local inertial frames attached to each space-time point, where the equations are fully Lorentz covariant and the propagation should be subluminal. Let us envisage a solution of the Maxwell-Chern-Simons equation in response to an external vector and/or axial current source placed \textit{on the boundary}. The AVV three point function extracted above links the source to the boundary value of the solution for the vector and axial vector field. The subluminality of the solution should be encoded in the analyticity with respect to complex momenta they depend on. On the other hand, a direct exploration of the relativistic causal structure of the AVV function with respect to $q_1$ and $q_2$ involves theory of multiple complex variables and appears difficult. See, e.g. Ref.\cite{Bu2016} for numerical evidences of two-point functions.

\section{The UV and IR Convergence}
\label{sec:uv-ir-convergence}

Let us recall the AVV three-point function from field theoretic perspectives. The power counting argument leads to the degree of UV divergence $1$, but the vector current conservation factors out two powers of external momenta, leaving the effective degree of divergence $-1$.  Indeed, the explicit calculation to the one-loop order gives rise to a finite result once the $U_V(1)$ gauge invariance is maintained through a proper regularization. Taking the Pauli-Villars regularization as an example, the regularized AVV three-point function to one-loop order remains finite in the limit of an infinite  regulator mass. The above power counting argument applies only to the skeleton diagram. To higher orders in coupling constant, UV divergence emerges via radiative corrections of the propagators and vertices inside the Feynman diagrams underlying the three point function. Upon the wave function and coupling constant renormalizations, the UV divergence is removed leaving the result depending on the renormalization scale, such as $\Lambda_{\rm QCD}$ for QCD. In addition to UV divergence, infrared divergence of a Yang-Mills theory at nonzero temperature grows with the order of diagrams and becomes out of control beyond a certain power of the coupling constant. Nonperturbative effect, such as the magnetic mass is expected to eliminate the IR divergence. Being a conformal field even at quantum level, $\mathcal{N}=4$ super Yang-Mills theory is expected to be UV finite and its gravity dual provides a nonperturbative approach of calculation. Therefore, both the UV and IR finiteness  should be reflected in the three-point function calculated via AdS/CFT correspondence and we shall prove below that this is indeed the case.  In this sense, our result also lends a support to the validity of the conjectured AdS/CFT correspondence. 

It follows from eqs. (\ref{J_perp}) and (\ref{J_0}) that to prove the UV/IR convergence amounts to prove the convergence of the following integrals
\begin{equation}
  I = \int_0^1 \ \mathrm{d}u\, (1-u^2) \bm{\mathcal{G}}_\perp(q_1,q_2|u) \psi_1(q|u)
  \label{integral1}
\end{equation}
and
\begin{equation}
  J = \int_0^1 \ \mathrm{d}u\, u(1-u^2)\mathcal{M}(q_1,q_2|u)\phi_1(q|u) \;,
  \label{integral2}
\end{equation}
with $q=q_1+q_2$, where $\psi_1(q|u)$ and $\phi_1(q|u)$ are the in-falling solutions of (\ref{eq:29}) and (\ref{eq:36}) normalized according to (\ref{normalization}), and $\bm{\mathcal{G}}_\perp(q_1,q_2|u)$ and $\mathcal{M}(q_1,q_2|u)$ are given by (\ref{eq:35}), (\ref{project}) and (\ref{eq:39}) that relate to $\psi_1$ and $\phi_1$ via (\ref{free}). The integration limits $u=0,1$ are the regular points of the Fuchs equations (\ref{eq:29}) and (\ref{eq:36}) with $u=0$ corresponding UV limit and $u=1$ to IR limit. Between them ($0<u<1$), the solutions are analytic. So are $\bm{\mathcal{G}}_\perp(q_1,q_2|u)$ and $\mathcal{M}(q_1,q_2|u)$. As long as the integrands are sufficiently well-behaved near the upper/lower limit, the integrals converge and our theme is proved.

\vspace{.5cm}
\noindent\underline{\bf{Lower Limit (UV)}}
\vspace{.5cm}

According to Table \ref{tab:indexes}, the asymptotic forms of $\psi_1(q|u)$ and $\phi_1^\prime(q|u)$ as $u\to 0$ are given by 
\begin{equation}
  \psi_1(q|u) = O(1) + O(u\ln u) \;; \qquad  
  \phi_1^\prime(q|u) = O(\ln u) \;.
  \label{logarithmic}
\end{equation}
It follows that,
\begin{equation}
  \bm{\mathcal{V}}(q|u)=O(1)  \;;\qquad  \bm{\mathcal{V}}'(q|u)=O(\ln u) \;,
  \label{V_UV}
\end{equation}
and
\begin{equation}
  \mathcal{A}_0'(q|u)=O(\ln u) \;;\qquad \mathcal{A}_0''(q|u)=O\left(\frac{1}{u}\right)
  \label{A_UV}
\end{equation}
in accordance with (\ref{free}). Consequently
\begin{equation}
\begin{aligned}
  \bm{\mathcal{B}}(q_1|u) &= 2i\pi\, \bm{\mathfrak{q}}_1\times\bm{\mathcal{V}}(q_1|u)=O(1) \;;\\
  \bm{\mathcal{E}}(q_1|u) &= 2i\pi \, \mathfrak{w}_1T\bm{\mathcal{V}}(q_1|u)=O(1) \;.
\end{aligned}
\label{other_UV}
\end{equation}
Substituting (\ref{V_UV}), (\ref{A_UV}) and (\ref{other_UV}) to RHS of (\ref{project}) and (\ref{eq:39}), we find that the integrands of (\ref{integral1}) and (\ref{integral2})
\begin{align}
  (1-u^2) \bm{\mathcal{G}}_\perp(q_1,q_2|u) &= O(\ln u) \;;\\
  u(1-u^2)\mathcal{M}(q_1,q_2|u)\phi_1(q|u) &= O(\ln u) \;.
\end{align}
So their singularities are not strong enough to give rise to UV divergence.

As a side remark, the logarithmic divergence of (\ref{logarithmic}) does show up in the first term of (\ref{J_perp}). But this divergence pertains to the zeroth power of $\kappa_{_\mathrm{CS}}$ and does not contribute to the chiral magnetic current. This UV divergence is the holographic version of the logarithmic divergence of the self-energy of $U_V(1)$ gauge boson in field theory and is cancelled by the holographic counter term $S_\text{c.t.}$ of eq.(\ref{eq:1}).

\vspace{.5cm}
\noindent\underline{\bf{Upper Limit (IR)}} 
\vspace{.5cm}

As $u \to 1^-$, the in-falling condition \eqref{normalization} implies the asymptotic forms:
\begin{align}
  \bm{\mathcal{V}}(q|u) &\sim (1-u)^{-i \frac{\mathfrak{w}}{2}} \;;\\
  \bm{\mathcal{V}}'(q_1|u) &\simeq \frac{i \, \mathfrak{w}_1}{2(1-u)}\bm{\mathcal{V}}_\perp(q_1|u)  \;;\\
  \mathcal{A}_0'(q|u) &\sim (1-u)^{-i\frac{\mathfrak{w}}{2}}   \;;\\
  \mathcal{A}_0''(q_2|u) &\simeq \frac{i \,\mathfrak{w}_2}{2(1-u)}\bm{\mathcal{A}}_0'(q_2|u)  \;;\\
  \mathcal{A}_\parallel'(q_2|u) &\simeq -\frac{\mathfrak{w}_2}{2|\bm{\mathfrak{q}}_2|(1-u)} \mathcal{A}_0'(q_2|u) \;.  
  \label{All_IR}
\end{align}
Substituting these asymptotic forms to RHS of (\ref{eq:35}) and (\ref{eq:39}), we find that the leading singularity of the order $\frac{1}{(1-u)^2}$ get cancelled, leaving
  \begin{eqnarray}
    \bm{\mathcal{G}}_\perp(q_1,q_2|u) \psi_1(q|u) &\sim& \frac{1}{1-u} \;;\\
    \mathcal{M}(q_1,q_2|u) &\sim& \frac{1}{1-u}  \;,
  \end{eqnarray}
  which make the integrals (\ref{integral1}) and (\ref{integral2}) convergent at the upper limit. The cancellation of the leading singularity in (\ref{eq:35}) follows from the relations
  \begin{align}
    \bm{\mathcal{B}}(q_1|u) = i\, 2\pi T \, \bm{\mathfrak{q}}_1 \times \bm{\mathcal{V}}(q_1|u)  \;;\qquad
    \bm{\mathcal{E}}(q_1|u) =  i \, 2 \pi T\,  \mathfrak{w}_1\bm{\mathcal{V}}(q_1|u)
  \end{align}
  and the cancellation in (\ref{eq:39}) follows from the observation that
  \begin{equation}
    \left[ u \mathcal{A}_0'(q_2|u) \; \big(\bm{\mathfrak{q}}_2 \cdot \bm{\mathcal{B}}(q_1|u) \big) \right]' 
    \simeq \frac{i\, \mathfrak{w}}{2(1-u)} \left[ u \mathcal{A}_0'(q_2|u) \; \big(\bm{\mathfrak{q}}_2 \cdot \bm{\mathcal{B}}(q_1|u) \big) \right] \;.
  \end{equation}

    Consequently, the chiral magnetic current and its induced charge driven by an external magnetic field and axial vector potential together with the response kernels $\Lambda_{ij0}(q_1,q_2)$ and $\Lambda_{0j0}(q_1,q_2)$ are free from UV and IR divergence and our theme is thereby proved.

    A curious divergence of the AVV three-point function at three loop level was discovered in the axial anomaly in the context of the massless QED at zero temperature \cite{Ansel'm}, where the two photons emerging from the AVV triangle diagram are re-scattered via a fermion loop.  As the vector and axial vector field in the bulk do not contribute to the internal lines of the Feynman diagrams of the super-Yang-Mills on the boundary, this complication does not arise here.

% %%%%%%%
\section{Concluding Remarks and Outlooks}
\label{sec:concluding}

In this work, we developed the holographic formulation of the chiral magnetic current for arbitrary energy-momenta of the external magnetic field $\bm{\mathsf{B}}(q_1)$ and temporal component of the axial-vector potential $\mathsf{A}_0(q_2)$ with the latter proxies the space-time variations of the chiral imbalance because of the topological fluctuations of QCD. The gauge theory on the AdS-boundary is the $\mathcal{N}=4$ super-Yang-Mills of large $N_c$ and strong `t Hooft coupling. $\bm{\mathsf{B}}(q_1)$ and $\mathsf{A}_0(q_2)$ come from the boundary values of the bulk vector and axial-vector potential, which correspond to gauged $U(1)$ subgroups of the global $U(4)$ R-symmetry of the super-Yang-Mills.  The kernel relating $\bm{\mathsf{B}}(q_1)$ and  $\mathsf{A}_0(q_2)$ to the vector current corresponds to the $(0ij)$ component of the AVV three-point function $\Delta_{\rho\mu\nu}(-q_1-q_2, q_1)$. For small but nonzero $q_2$, the chiral magnetic response turns out to be non-local because of $D(q_2)=\mathfrak{w}_2- |\bm{\mathfrak{q}}_2|^2+...$ in the denominator and we replicated the field theoretic result regarding the sensitivity of the three-point function to the order of infrared limit $q_2\to 0$ when $q_1$ is also small \cite{Hou,Miklos}.  For arbitrary momenta, the kernel can be expressed in terms of two Heun functions that are difficult to manipulate because of their complexity.  Finally we proved that the AVV function does not suffer from UV and IR divergences, resonating with the finiteness of the super-Yang-Mills on the boundary. 
    
    The case of a homogeneous chiral imbalance requires special handling because of the additional integration constant incurred in the 0-th order solution for the bulk axial vector potential component $\mathcal{A}_0(q|u)$ at $\bm{q}=0$. To reproduce the classical CME formula (\ref{CME}), we follow the gauge invariant definition of the axial chemical potential in \cite{Gynther2011} by setting $\mathcal{A}_0=0$ on the boundary at cost of introducing a nonzero $\mathcal{A}_0(0|u)$ at the horizon, a singular field configuration in the local inertial frame there. Though the singularity has no known physical impact, the issue reflects the difficulty of defining the axial chemical potential associated to a non-conserved axial charge \cite{Rubakov2010}. Nevertheless, this recipe generates the first term of the chiral magnetic current \eqref{eq:21} which restores (\ref{CME}) as its special case with a constant magnetic field. A by-product of our formulation is an analytic expression for this part of the chiral magnetic current in terms of an ordinary hypergeometric function for a homogeneous but time-dependent magnetic field and this type of magnetic field was assumed in some hydrodynamic simulation of CME in RHIC.

    There is a vast amount of literature on the holographic chiral magnetic effect, and the probe limit is frequently employed there, which facilitates the expansion to nonlinear orders of external vector and/or axial vector electromagnetic field without considering the metric fluctuations \cite{Gynther2011,Bu2016a,Bu2016,Bu2019}. The chemical potential is modeled as the background temporal component of the vector and/or axial vector potential and the fluctuation away from the background is assumed small, 
    \begin{align}
      |\mu_A| \gg |\mathsf{A}_0(x)|  \label{eq:12}
    \end{align}
with $\mathsf{A}_0(x)=\int\frac{\mathrm{d}^4q}{(2\pi)^4}e^{iq\cdot x}\mathsf{A}_0(q)$. The transport coefficients thus obtained carry all powers of $\mu_A$. What follows from the power counting argument in section~\ref{sec:action} is that we do not have to assume the probe limit as far as AVV three-point function is concerned where $\mu_A$ is treated as a part of axial vector field. This is attributed to the fact that the stress tensor contributing to the metric fluctuations starts with quadratic power in the vector or axial vector fields. In case of the $\mathcal{N}=4$ super-Yang-Mills on the boundary, the coefficients on RHS of the Einstein equation (\ref{eq:1_ren}), $\kappa_{\text{M}}/\kappa_{\text{EH}} = \mathcal{O}(1)$ and is thereby not tunable. To higher powers beyond  bi-linear terms in $\bm{\mathsf{V}}(q_1)$,  $\mathsf{A}_0(q_2)$, the metric fluctuations have to be brought in and the probe limit is not applicable. Therefore, the coefficient of the current beyond the linear power in $\mu_A$ evaluated under the probe limit is no longer pertaining to the $\mathcal{N}=4$ super-Yang-Mills. Nevertheless, the chiral magnetic conductivity evaluated under the probe limit and the condition (\ref{eq:12}), e. g. References\cite{Gynther2011,Bu2016a}, contains AVV contribution to the leading power in $\mu_A$, one of the independent momenta, $q_2$, is set to zero. What we developed here is the formulation of the three-point function with both $q_2$ and $q_1$ nonzero and thereby displaying the non-local response and non-trivial IR limit reflected in the formulas (\ref{J_spatial}) and (\ref{charge_0}) for small momenta, which appears more realistic from the perspective of dynamic process of the axial charge creation/annihilation via the topological transitions of QCD that accompanies the development of the QGP fireballs in heavy ion collisions.
    
In section \ref{sec:smallq}, we have presented preliminary charts of the AVV contribution of the chiral magnetic current for low momenta $q_1$ and $q_2$ which shows that a spacetime dependent $\mathsf{A}_0$ can generate a sizable impact on the chiral magnetic signal.  Going forward, we would like to explore the AVV contribution over the entire momentum space of $q_1$ and $q_2$ and its phenomenological implication. Unlike a two point function a 3D plot of which suffices, a visual illustration of a three point function is far more challenging since we have here five rotational invariant quantities made of $q_1$ and $q_2$ ( $\omega_{1,2}$, $\bm{q}_{1,2}^2$ and $\bm{q}_1 \cdot \bm{q}_2$) to explore.  An alternative approach is to convolute the AVV three point functions with appropriate profiles of magnetic field and $\mathsf{A}_0$ to simulate the electric current of a single collision event in coordinate space. For example, we may model the axial charge creation/annihilation by the following stochastic process 
\begin{equation}
  \left(\frac{\partial}{\partial t}-D\nabla^2+\frac{1}{\tau}\right)n_A=g(x)
\end{equation}
where $D$ and $\tau$ are phenomenological diffusion constant and lifetime, and $g(x)$ is a white noise describing QCD topological transitions. A stochastic profiles $\mathsf{A}_0$ will be generated through a phenomenological susceptibility. The current-current correlation that survives the stochastic average and summation over all collision events
\begin{align}
  \langle J_i^{\rm CME}(x)J_j^{\rm CME}(x') \rangle = \langle \mathsf{J}_i(x)\mathsf{J}_j(x') \rangle + \langle J_i^{\rm AVV}(x)J_j^{\rm AVV}(x') \rangle
\end{align}
carries the CME signals to be detected and the AVV term gives rise to a new contribution beyond the hydrodynamic simulations in the literature. Moreover, different choices of the parameters $D$ and $\tau$ as well as the profile of the magnetic field would highlight different regions of the momenta $q_1$ and $q_2$.

Our formulation can be readily generalized to explore the chiral separation effect (CSE) under a magnetic field and a space-time dependent chemical potential which is conjugate to the charges associated with the vector potential. As the vector current is conserved, the special treatment, such as \eqref{eq:57}, of the homogeneous component of the $\mathcal{V}_0$ in the bulk may not be warranted and the ambiguity associated with the IR limit may disappear. We hope to report our progress in this direction in near future.
%    %%%%%%%%%%%%%%%%%%%%%%%%%%%%%%%%%%%%%%%%%%%%%%%%%% 

\acknowledgments
L.Yin is supported by Guangdong Major Project of Basic and Applied Basic Research No. 2020B0301030008 and Science and Technology Program of Guangzhou No. 2019050001. D-f. Hou and H-c. Ren are supported in part by the NSFC Grant Nos. 11735007, 11890711, 11890710.

% %%%%%%

%\newpage
\appendix
% %%%
\section{Notation List}
\label{sec:notation-list}
\begin{tabular}{l|p{12cm}}
  $\diamond$\; $\mu_A$ & The axial chemical potential, defined in eq.~\eqref{eq:57}.  \vspace{.15cm}\\

  $\diamond$\;  $\bm{\mathsf{B}}(q)$ &  The magnetic field participating in the CME current, see \eqref{eq:21} and \eqref{total_current}. \vspace{.15cm}\\
  $\diamond$\; $\mathsf{J}^\mu(q)$ &  The CME current because of the chiral chemical potential $\mu_A$, see \eqref{eq:52}, $\mathsf{J}^\mu(q) = \{0, \bm{\mathsf{J}}(q)\}$, the spatial component is derived in eq.~\eqref{CME_general}. \vspace{.15cm}\\
  $\diamond$\; $\mathcal{J}^\mu(q_1,q_2)$ & The CME current because of the AVV three-point function, see the convolution in \eqref{eq:58}, the two 4-momenta $q_1$ and $q_2$ reflect the spacetime-dependence of magnetic field and chiral imbalance, respectively. \vspace{.15cm}\\

  $\diamond$\; $\mathrm{F}^{V,A}_{MN}, \mathrm{A}_M, \mathrm{V}_M$  & The gauge field tensors and potentials in the bulk theory, see the classical action~\eqref{eq:2} and the EoM \eqref{eq:6}\eqref{eq:7}\eqref{eq:1_ren}.  \vspace{.15cm}\\

  $\diamond$\; $\mathbb{F}^{V,A}_{MN}, \mathbb{A}_M, \mathbb{V}_M$  & The fluctuations of vector/axial gauge fields in the bulk theory~\eqref{eq:60}, their EoM are the inhomogeneous Maxwell equations~\eqref{eq:6_ren}\eqref{eq:7_ren}.   \vspace{.15cm}\\

  $\diamond$\;  $\mathsf{A}_0, \bm{\mathsf{V}}$  & The non-zero components of axial/vector gauge potentials $\mathbb{A}_\mu, \mathbb{V}_\mu$  on AdS-boundary, for coordinate representation see eq.~\eqref{eq:boundary}, for Fourier representation eq.~\eqref{eq:boundary1}. \vspace{.15cm}\\

  $\diamond$\; $\mathcal{F}^{V,A}_{MN}, \mathcal{A}_M, \mathcal{V}_M$  & The fluctuations of vector/axial gauge tensor and gauge potentials in bulk to the zeroth order of $\kappa_{_\text{CS}}$~\eqref{eq:53_1}\eqref{eq:53_2}, their EoM are the homogeneous Maxwell equations~\eqref{eq:6_ren1}\eqref{eq:7_ren1}. The temporal component $\mathcal{A}_0(q|u)$ is related to the axial chemical potential in eq.~\eqref{eq:57}; the spatial component $\bm{\mathcal{V}}$ produces the electromagnetic field $\bm{\mathsf{B}}(q); \bm{\mathsf{E}}(q)$ from eq.~\eqref{free} and eq.~\eqref{eq:30}. \vspace{.15cm}\\

  $\diamond$\; $\mathtt{F}^{V,A}_{MN}, \mathtt{A}_M, \mathtt{V}_M$  & The fluctuations to the first order CS coupling $\kappa_{_\text{CS}}$, that are the solutions from the inhomogeneous Maxwell equations~\eqref{eq:6_ren2}~\eqref{eq:7_ren2} by first iteration. \vspace{.15cm}\\

  $\diamond$\; $\psi_1; \psi_2$   &  The two linearly-independent solutions of $\bm{\varPsi} = \{ \bm{\mathcal{V}}_\perp, \bm{\mathcal{A}}_\perp \}$ \eqref{eq:29}, and $\psi_1$ refers to the falling wave solution, see the normalization condition in eq.~\eqref{normalization} , they play a key role in solving the longitudinal component of the fluctuation $\mathsf{V}_\perp$ eq.~\eqref{eq:43_perp}.  \vspace{.15cm}\\

  $\diamond$\; $\phi_1; \phi_2$   & The two linearly-independent solutions of $\bm{\varPhi} = \{ \mathcal{V}_0', \mathcal{A}_0' \}$ \eqref{eq:36}, they play a key role in solving the temporal component of the fluctuation $\mathsf{V}_0'$ eq.~\eqref{eq:43_0}. 
\end{tabular}

% %%%%%
\section{Inhomogeneous Maxwell Equation in Coordinate and Fourier Space}
\label{sec:inhom-maxw-equat}

Substituting the AdS-Schwarzschild metric (\ref{eq:16}) and the gauge condition $\mathtt{V}_u=\mathtt{A}_5=0=\mathcal{V}_u=\mathcal{A}_5=0$ into (\ref{eq:6_ren2}) and (\ref{eq:7_ren2}),  we obtain explicit coordinate representation of the field equations up to the linear order in $\kappa_{_\text{CS}}$, i.e.

\begin{align}
  \partial_0 \partial_5\mathtt{V}_0 - f \, \partial_k \partial_5\mathtt{V}_k &= - \frac{3}{2 \sqrt{2}(\pi T)^2L}\frac{\kappa_{_\text{CS}}}{\kappa_\text{M}}  \varepsilon^{kij} \big[ \mathcal{F}_{0k}^A \mathcal{F}_{ij}^V + \mathcal{F}_{0k}^V \mathcal{F}_{ij}^A \big]  \\
   \partial_0 \partial_5\mathtt{A}_0 - f \, \partial_k \partial_5\mathtt{A}_k &= - \frac{3}{2 \sqrt{2}(\pi T)^2L}\frac{\kappa_{_\text{CS}}}{\kappa_\text{M}}\varepsilon^{kij} \big[ \mathcal{F}_{0k}^V \mathcal{F}_{ij}^V + \mathcal{F}_{0k}^A \mathcal{F}_{ij}^A \big] 
\end{align}

\begin{align}
 \partial_5^2 \mathtt{V}_0 - \frac{1}{(2\pi T)^2 uf}\, \partial_k \big[\partial_k \mathtt{V}_0 - \partial_0 \mathtt{V}_k \big] &= - \frac{3}{2 \sqrt{2}(\pi T)^2L}\frac{\kappa_{_\text{CS}}}{\kappa_\text{M}}  \varepsilon^{kij} \big[ \partial_5\mathcal{A}_k \mathcal{F}_{ij}^V + \partial_5\mathcal{V}_k  \mathcal{F}_{ij}^A \big] \\
 \partial_5^2\mathtt{A}_0 - \frac{1}{(2\pi T)^2 uf}\, \partial_k \big[\partial_k \mathtt{A}_0 - \partial_0 \mathtt{A}_k \big] &=
  - \frac{3}{2 \sqrt{2}(\pi T)^2L}\frac{\kappa_{_\text{CS}}}{\kappa_\text{M}} \varepsilon^{kij} \big[ \partial_5\mathcal{V}_k \mathcal{F}_{ij}^V + \partial_5\mathcal{A}_k \mathcal{F}_{ij}^A \big] 
\end{align}

\begin{equation}
\begin{aligned}
  \partial_5 \big[ f \partial_5\mathtt{V}_k \big]  & - \frac{ \partial_0 \big[\partial_0 \mathtt{V}_k  - \partial_k \mathtt{V}_0 \big] }{ (2\pi T)^2\, uf }  + \frac{ \partial_l[\partial_l \mathtt{V}_k - \partial_k \mathtt{V}_l]}{ (2\pi T)^2\, u}   = \\
 &  -  \frac{3}{2 \sqrt{2}(\pi T)^2L}\frac{\kappa_{_\text{CS}}}{\kappa_\text{M}} \varepsilon^{kij} \big[  \big( \partial_5\mathcal{V}_0 \mathcal{F}_{ij}^A +  \partial_5\mathcal{A}_0 \mathcal{F}_{ij}^V \big) - 2 \big( \partial_5\mathcal{V}_i \mathcal{F}_{0j}^A + \partial_5\mathcal{A}_i \mathcal{F}_{0j}^V \big) \big]
\end{aligned}
\end{equation}
\begin{equation}
\begin{aligned}
  \partial_5\big[ f \partial_5\mathtt{A}_k \big]  & - \frac{ \partial_0 \big[\partial_0 \mathtt{A}_k - \partial_k \mathtt{A}_0 \big] }{ (2\pi T)^2\, uf } +  \frac{ \partial_l[\partial_l \mathtt{A}_k - \partial_k \mathtt{A}_l]}{ (2\pi T)^2\, u} = \\
  & - \frac{3}{2 \sqrt{2}(\pi T)^2L}\frac{\kappa_{_\text{CS}}}{\kappa_\text{M}} \varepsilon^{kij} \big[  (\partial_5\mathcal{A}_0 \mathcal{F}_{ij}^A + \mathcal{F}_{ij}^V \partial_5\mathcal{V}_0) - 2 (\partial_5\mathcal{A}_i \mathcal{F}_{0j}^A + \mathcal{F}_{0j}^V \partial_5\mathcal{V}_i) \big]
\end{aligned}
\end{equation}
where
$\partial_5 \mathtt{V}_\mu \equiv \frac{\partial \mathtt{V}_\mu}{\partial u}$ and we have separated the time index ``0'' and the AdS radial index ``5'' from the spatial indices on the boundary.

Making Fourier transformation with respect to the boundary coordinate $x^\mu$ on both sides with
\begin{equation}
\partial_\mu \to iq_\mu=i(-\omega, \bm{q})
\end{equation} 
we find that

\begin{align}
  \mathfrak{w}\mathtt{A}_0' + f \big(\bm{\mathfrak{q}} \cdot \bm{\mathtt{A}}\big)' &= \frac{\kappa_{_\text{CS}}}{\kappa_\text{M}}G^5_A(q|u) \\
  \mathfrak{w}\mathtt{V}_0' + f \big( \bm{\mathfrak{q}} \cdot \bm{\mathtt{V}}\big)' &= \frac{\kappa_{_\text{CS}}}{\kappa_\text{M}}G^5_V(q|u)
\end{align}

\begin{align}
  \mathtt{A}_0'' - \frac{1}{uf} \big[ |\bm{\mathfrak{q}}|^2 \mathtt{A}_0 + \mathfrak{w} ( \bm{\mathfrak{q}} \cdot \bm{\mathtt{A}}) \big] &= \frac{\kappa_{_\text{CS}}}{\kappa_\text{M}}G^0_A(q|u) \\
  \mathtt{V}_0'' - \frac{1}{uf} \big[|\bm{\mathfrak{q}}|^2 \mathtt{V}_0 + \mathfrak{w} (\mathfrak{q} \cdot \bm{\mathtt{V}}) \big] &= \frac{\kappa_{_\text{CS}}}{\kappa_\text{M}}G^0_A(q|u)
\end{align}

\begin{align}
  \mathtt{A}_k'' + \frac{f'}{f} \mathtt{A}_k' + \frac{1}{uf^2} \big[\mathfrak{w}^2 \mathtt{A}_k + \mathfrak{w} \mathfrak{q}_k \mathtt{A}_0\big] - \frac{1}{u f} \big[ |\bm{\mathfrak{q}}|^2 \mathtt{A}_k - \mathfrak{q}_k (\bm{\mathfrak{q}} \cdot \bm{\mathtt{A}})\big] &= \frac{\kappa_{_\text{CS}}}{\kappa_\text{M}}G_A^k(q|u) \\
  \mathtt{V}_k'' + \frac{f'}{f} \mathtt{V}_k' + \frac{1}{uf^2} \big[\mathfrak{w}^2 \mathtt{V}_k + \mathfrak{w} \mathfrak{q}_k \mathtt{V}_0\big] - \frac{1}{u f} \big[ |\bm{\mathfrak{q}}|^2 \mathtt{V}_k - \mathfrak{q}_k (\bm{\mathfrak{q}} \cdot \bm{\mathtt{V}})\big] &= \frac{\kappa_{_\text{CS}}}{\kappa_\text{M}}G_V^k(q|u)
\end{align}

where we have introduced dimensionless 4-momenta via \eqref{eq:dimensionless} and denoted the derivative with respect to $u$ by a prime for brevity.
 The explicit form of the inhomogeneous terms $G^5_V(q|u)$, $G^0_V(q|u)$ and $G_V^k(q|u)$ are displayed in (\ref{eq:25}) as a convolution with the explicit form of the integrand given by \eqref{eq:31}, \eqref{eq:33} and (\ref{eq:35}). The parallel expressions of 
$G^5_V(q|u)$, $G^0_V(q|u)$ and $G_V^k(q|u)$, which are not used in this work, are summarized in the following convolution form.

\begin{align}
  G_{_\mathtt{A}}^M(q|u) = \int \frac{\mathrm{d}^4q_1}{(2\pi)^4} \frac{\mathrm{d}^4q_2}{(2\pi)^4} (2\pi)^4 \delta^4(q_1+q_2-q)\; \mathcal{G}_{_\mathrm{A}}^M(q_1,q_2|u) \equiv \int_{q_1,q_2} \, \mathcal{G}_{_\mathrm{V}}^\mu(q_1,q_2|u)  \;,
  \label{eq:25_axial}
\end{align}
where the integrand
\begin{align}
  \mathcal{G}_{_\mathrm{A}}^5(p,l|u) &= - \frac{3}{2\sqrt{2}} \frac{p_0}{(\pi T)^3L}\, \bm{\mathcal{V}}_\perp(p|u) \cdot \bm{\mathcal{B}}(p|u)  \label{eq:30_axial}\\
  \mathcal{G}_{_\mathrm{A}}^0(p,l|u) &= - \frac{3}{2\sqrt{2}} \frac{1}{(\pi T)^2L}\; \bm{\mathcal{V}}_\perp(p|u) \cdot \bm{\mathcal{B}}(p|u)   \label{eq:32_axial} \\
  \mathcal{G}_k^A(p,l|u) &= - \frac{3}{\sqrt{2}} \frac{1}{(\pi T)^2 L f} \left[ \mathcal{V}_0'(p|u) \mathcal{B}_k(p|u) - \big(\bm{\mathcal{E}} \times \bm{\mathcal{V}}_\perp'\big)_k(p|u) \right]   \label{eq:34_axial}  \;,
\end{align}
where, for $u \to 0 $, the classical fields $\bm{\mathcal{B}}(p|u)$ and $\bm{\mathcal{E}}(p|u)$ become the magnetic field and electric field on the boundary according to (\ref{eq:30}).

\section{Zeroth Order Solution in Terms of Heun Functions}
\label{sec:heun}

Each of the differential equations (\ref{eq:29}) and (\ref{eq:36}) is a Fuchs equation with four regular points at $u=0,1,-1,\infty$. Making the transformation
\begin{equation}
  \begin{aligned}
  \Psi &=& (1-u)^{-i\frac{\mathfrak{w}}{2}}\left(\frac{1+u}{2}\right)^{\frac{\mathfrak{w}}{2}}f_{\rm I}  \\
  \Phi &=& (1-u)^{-i\frac{\mathfrak{w}}{2}}\left(\frac{1+u}{2}\right)^{\frac{\mathfrak{w}}{2}}f_{\rm II}
\end{aligned}
\end{equation}
and $z=\frac{1-u}{2}$, $f_s$ with $s={\rm I,II}$ satisfy the standard Heun equation
\begin{equation}
  z(z-1)(z-a_s)\frac{d^2f_s}{dz^2}+[(\alpha_s+\beta_s+1)z^2-[\alpha_s+\beta_s+1-\delta_s+(\gamma_s+\delta_s)a_s]z+a_s\gamma_s]\frac{df_s}{dz}+\alpha_s\beta_s(z-b_s)f_s=0
\end{equation}
where the parameters 
\begin{equation}
\begin{aligned}
  a_{\rm I} &= a_{\rm II} = \frac{1}{2} \\ 
  \alpha_{\rm I} &= \alpha_{\rm II} = \frac{1}{2}(1-i)\mathfrak{w} \\
  \gamma_{\rm I} &= \gamma_{\rm II} = 1-i\mathfrak{w} \\
  \delta_{\rm I} &= \delta_{\rm II} = 1+\mathfrak{w}  \\
  \beta_{\rm I} &= \frac{1}{2}(1-i)\mathfrak{w}+1 \\
  \beta_{\rm II} &= \frac{1}{2}(1-i)\mathfrak{w}+2 \\    
  b_{\rm I} &= \frac{\mathfrak{w}^2 + \left(-\frac{1}{2}+i\right)\mathfrak{w}- |\bm{\mathfrak{q}}|^2}{i\mathfrak{w}^2-(1-i)\mathfrak{w}} \\
  b_{\rm II} &= \frac{\left(1-\frac{i}{2}\right)\mathfrak{w}^2 + \left(-1+\frac{3}{2}i\right)\mathfrak{w}-|\bm{\mathfrak{q}}|^2}{i\mathfrak{w}^2-2(1-i)\mathfrak{w}}
\end{aligned}
\end{equation}
The indices at the regular points $z=0,1,a_s,\infty$ are $(0,1-\gamma_s)$, $(0,1-\delta_s)$, $(0,1-\epsilon_s)$ and $(\alpha_s,\beta_s)$ with 
$\epsilon_s=\alpha_s+\beta_s-\gamma_s-\delta_s+1$.
In terms of the standard notation of the Heun function in \cite{Whittaker}, the in-falling solutions normalized by the conditions (\ref{normalization}) are given by (\ref{normalization}):
\begin{equation}
\begin{aligned}
  \psi_1(q|u) &= (1-u)^{-i\frac{\mathfrak{w}}{2}}\left(\frac{1+u}{2}\right)^{\frac{\mathfrak{w}}{2}}F\left(a_{\rm I},b_{\rm I};\alpha_{\rm I},\beta_{\rm I},\gamma_{\rm I},\delta_{\rm I};\frac{1-u}{2}\right) \\
  \phi_1(q|u) &=(1-u)^{-i\frac{\mathfrak{w}}{2}}\left(\frac{1+u}{2}\right)^{\frac{\mathfrak{w}}{2}}F\left(a_{\rm II},b_{\rm II};\alpha_{\rm II},\beta_{\rm II},\gamma_{\rm II},\delta_{\rm II};\frac{1-u}{2}\right)
\end{aligned}
\end{equation}
Interestingly, the infinity of (\ref{eq:29}) becomes an ordinary point in the homogeneous limit, $\bm{\mathfrak{q}} \to 0 $, and the Heun equation is reduced to a hypergeometric equation with the in-falling solution
\begin{equation}
  \psi_1(q|u) = \left(\frac{1-u}{1+u}\right)^{-i\frac{\mathfrak{w}}{2}}F\left(\frac{1-i}{2}\mathfrak{w},-\frac{1+i}{2}\mathfrak{w};1-i\mathfrak{w};\frac{1-u}{1+u}\right)
\end{equation}
and
\begin{equation}
  \psi_1(q|0) = \frac{\Gamma \left( 1-i\mathfrak{w}\right)}{\Gamma\left(\frac{1-i\mathfrak{w}}{2}\right)\Gamma \left( \frac{3-i\mathfrak{w}}{2}\right)} \;.
\end{equation}  

\section{Special Solutions by Variation of Parameter }
\label{sec:spec-solut-vari}

It follows from the method of variation of parameters that the general solution of an inhomogeneous 2nd-order differential equation $L_u \Psi(u) = g(u)$ with differential operator
\begin{align}
  L_u := \frac{\mathrm{d}^2}{\mathrm{d}u^2} +p(u) \frac{\mathrm{d}  }{\mathrm{d}u} +q(u)
\end{align}
is given by
\begin{align}
  \Psi(u) = c_1 \psi_1(u) + c_2 \psi_2(u) + \psi_1(u) \int_u^1 \ \mathrm{d}\xi \frac{g(\xi)}{W(\xi)} \psi_2(\xi) - \psi_2(u) \int_u^1 \ \mathrm{d}\xi \frac{g(\xi)}{W(\xi)}\psi_1(\xi) \;.
\end{align}
where $\psi_1(u)$ and $\psi_2(u)$ are the two linearly independent solution of the homogeneous equation $L_u\psi(u) =0$ and $W(u)$ is their Wronskian. The constants $c_1$ and $c_2$ are determined by appropriate boundary conditions. 

For the inhomogeneous equations (\ref{V_perp}) and (\ref{eq:37}) for $\bm{\mathtt{V}}_\perp$ and $V_0'$ with in-falling and outgoing solutions of the homogeneous equation, $\psi_1(u)$ and $\psi_2(u)$, at $u=1$, the in-falling condition of $\Psi(u)$ there set $c_2=0$. The Dirichlet like boundary condition at $u=0$ for $\bm{\mathtt{V}}_\perp$ gives rise to the constant
\begin{align}
  c_1 = \frac{\Psi(0)}{\psi_1(0)} - \int_0^1 \ \mathrm{d}\xi \frac{g(\xi)}{W(\xi)} \psi_2(\xi) + \frac{\psi_2(0)}{\psi_1(0)} \int_0^1 \ \mathrm{d}\xi \frac{g(\xi)}{W(\xi)}\psi_1(\xi) \;,
\end{align}
and the solution (\ref{eq:43_perp}) together with (\ref{eq:40}) follow then. The Newman like boundary condition for $V_0'$, $uV_0''(u)\to 0$ as $u\to 0$ determines that 
\begin{align}
  c_1 =  - \int_0^1 \ \mathrm{d}\xi \frac{g(\xi)}{W(\xi)} \psi_2(\xi) + \frac{\psi_2'(0)}{\psi_1'(0)} \int_0^1 \ \mathrm{d}\xi \frac{g(\xi)}{W(\xi)} \psi_1(\xi) \;,  
\end{align}
and gives rise to the solution (\ref{eq:43_0}) together with (\ref{eq:43}).

\section{Low Momentum Expansion of the Diffusion Denominator}
\label{sec:diffusion}

To derive the low momentum expansion of the diffusion denominator $D(q)$, eq. (\ref{Dq}), we convert  the differential equation (\ref{eq:46}) into an integral equation via the method of variation of parameters, subject to the boundary condition of $F=1$ at $u=1$. We have
\begin{equation}
F = 1+\frac{1}{2}\int_u^1d\xi\xi(1-\xi^2)E\ln\frac{\xi^2}{1-\xi^2}-\frac{1}{2}\ln\frac{u^2}{1-u^2}\int_u^1d\xi\xi(1-\xi^2)E
\label{integraleq}
\end{equation}
where
\begin{equation}
E = -\frac{i\mathfrak{w}}{1-u}F' - \frac{i\mathfrak{w}(1+2u)}{2u(1-u^2)} F- \frac{\mathfrak{w}^2(4+3u+u^2)}{4u(1+u)(1-u^2)} F + \frac{|\bm{\mathfrak{q}}|^2}{u(1-u^2)}F
\end{equation}
As $F\simeq D(q)\ln u$ as $u\to 0$, both integrals on RHS of (\ref{integraleq}) are convergent and the logarithmic behavior comes from the third term, i.e.
\begin{equation}
  D(q) = -\int_0^1 \ \mathrm{d}u\ u(1-u^2)E
\end{equation}
The integral equation can be solved iteratively. Substituting the zeroth order solution $F=1$ into $E$, we find the leading order $D(q) = i\mathfrak{w}-|\bm{\mathfrak{q}}|^2$. Substituting the first order solution \eqref{eq:47} into $E$, we end up with the expansion of $D(q)$ to the next order, i. e. eq.\eqref{Dq}.

%%%%%%%%%%%%%%%%%%%%%%%%%%%%%%%%%%%%%%%%%%%%%%%%%%%%%%%%%%%%%
%\bibliographystyle{apsrev4-2}
%\bibliography{$HOME/Documents/ResearchBox/library} %

\begin{thebibliography}{9}


\bibitem{CME2} Kenji Fukushima, Dmitri E. Kharzeev, and Harmen J. Warringa, \textit{Chiral magnetic effect}, \href{https://journals.aps.org/prd/abstract/10.1103/PhysRevD.78.074033}{Phys. Rev. D {\bf 78}, 074033 (2008)}, [arXiv:0808.3382].
  
\bibitem{CME1} Dmitri E. Kharzeev, Larry D. McLerran, Harmen J. Warringa, \textit{The effects of topological charge change in heavy ion collisions: 'Event by event P and CP violation'}, \href{https://www.sciencedirect.com/science/article/abs/pii/S037594740800078X}{Nucl. Phys. A {\bf 803}, 227 (2008)}, [arXiv:0711.0950].

\bibitem{CME3} Dmitri E. Kharzeev, Harmen J. Warringa, \textit{Chiral magnetic conductivity}, \href{https://journals.aps.org/prd/abstract/10.1103/PhysRevD.80.034028}{Phys. Rev. D {\bf 80}, 034028 (2009)}, [arXiv:0907.5007].

\bibitem{CVE} Dam T. Son, Piotr Surowka,  \textit{Hydrodynamics with Triangle Anomalies}, \href{https://journals.aps.org/prl/abstract/10.1103/PhysRevLett.103.191601}{Phys. Rev. Lett. {\bf 103}, 191601 (2009)}, [arXiv:0906.5044].

\bibitem{CMEinSTAR} STAR Collaboration, \textit{Azimuthal charged-particle correlations and possible local strong parity violation}, \href{https://journals.aps.org/prl/abstract/10.1103/PhysRevLett.103.251601}{Phys. Rev. Lett. {\bf 103}, 251601 (2009)}, [arXiv:0909.1739];
  \newline \textit{Observation of charge-dependent azimuthal correlations and possible local strong parity violation in heavy ion collisions}, \href{https://journals.aps.org/prc/abstract/10.1103/PhysRevC.81.054908}{Phys. Rev. C {\bf 81}, 054908 (2010)}, [arXiv:0909.1717];
  \newline \textit{Fluctuations of charge separation perpendicular to the event plane and local parity violation in $\sqrt{s_{NN}}=200GeV$ Au+Au collisions at the BNL Relativistic Heavy Ion Collider}, \href{https://journals.aps.org/prc/abstract/10.1103/PhysRevC.88.064911}{Phys. Rev. C {\bf 88}, 064911 (2013)}, [arXiv:1302.3802];
  \newline \textit{Measurement of charge multiplicity asymmetry correlations in high-energy nucleus-nucleus collisions at $\sqrt{s_{NN}}=200GeV$}, \href{https://journals.aps.org/prc/abstract/10.1103/PhysRevC.89.044908}{Phys. Rev. C {\bf 89}, 044908 (2014)}, [arXiv:1303.0901];
  \newline \textit{Beam-energy dependence of charge separation along the magnetic field in Au+Au collisions at RHIC}, \href{https://link.aps.org/doi/10.1103/PhysRevLett.113.052302}{Phys. Rev. Lett. {\bf 113}, 052302 (2014)}, [arXiv:1404.1433].

\bibitem{CMEinALICE} ALICE Collaboration, \textit{Charge separation relative to the reaction plane in Pb-Pb collisions at $\sqrt{s_{NN}}=2.76TeV$}, \href{https://journals.aps.org/prl/abstract/10.1103/PhysRevLett.110.012301}{Phys. Rev. Lett. {\bf 110}, 012301 (2013)}, [arXiv:1207.0900];
  \newline \textit{Constraining the magnitude of the chiral magnetic effect with event shape engineering in Pb-Pb collisions at $\sqrt{s_{NN}}=2.76TeV$}, \href{https://doi.org/10.1016/j.physletb.2017.12.021}{Phys. Lett. {\bf B 777}, 151 (2018)}, [arXiv:1709.04723].

\bibitem{CMEinCMS} CMS Collaboration, \textit{Observation of charge-dependent azimuthal correlations in p-Pb collisions and its implication for the search for the chiral magnetic effect}, \href{https://journals.aps.org/prl/abstract/10.1103/PhysRevLett.118.122301}{Phys. Rev. Lett. {\bf 118} 122301 (2017) }, [arXiv:1610.00263];
  \newline \textit{Constraints on the chiral magnetic effect using charge-dependent azimuthal correlations in pPb and PbPb collisions at the CERN Large Hadron Collider}, \href{https://journals.aps.org/prc/abstract/10.1103/PhysRevC.97.044912}{Phys. Rev. C {\bf 97} 044912 (2018)}, [arXiv:1708.01602].

\bibitem{Son_Spivak} D. T. Son, and B. Z. Spivak, \textit{Chiral anomaly and classical negative magnetoresistance of Weyl metals}, \href{https://journals.aps.org/prb/abstract/10.1103/PhysRevB.88.104412}{Phys. Rev. B {\bf 88}, 104412 (2013)}, [arXiv:1206.1627]. 

\bibitem{CMEinSemimetals_1} Qiang Li, Dmitri E. Kharzeev, Cheng Zhang, Yuan Huang, I. Pletikosic, A. V. Fedorov, R. D. Zhong, J. A. Schneeloch, G. D. Gu, T. Valla, \textit{Chiral magnetic effect in ZrTe5}, \href{https://doi.org/10.1038/nphys3648}{Nat. Phys. {\bf 12}, 550 (2016)}, [arXiv:1412.6543].

\bibitem{CMEinSemimetals_2} Xiaochun, Huang et al., \textit{Observation of the Chiral-Anomaly- Induced Negative Magnetoresistance in 3D Weyl Semi-metal TaAs}, \href{https://journals.aps.org/prx/abstract/10.1103/PhysRevX.5.031023}{Phys. Rev. X 5, 031023 (2015)}, [arXiv:1503.01304].

\bibitem{Katsuhisa_Taguchi} Katsuhisa Taguchi, Yukio Tanaka, \textit{Axial current driven by magnetization dynamics in Dirac semimetals}, \href{https://journals.aps.org/prb/abstract/10.1103/PhysRevB.91.054422}{Phys. Rev. B {\bf 91}, 054422 (2015)}, [arXiv:1406.4636].

\bibitem{Hou} Defu Hou, Hui Liu, Hai-cang Ren, \textit{Some field theoretic issues regarding the chiral magnetic effect}, \href{https://doi.org/10.1007/JHEP05(2011)046}{J. High Energy Phys. {\bf 2011}, 46 (2011)}, [arXiv:1103.2035].

\bibitem{JHGao} Jian-Hua Gao, Zuo-Tang Liang, Shi Pu, Qun Wang, Xin-Nian Wang, \textit{Chiral Anomaly and Local Polarization Effect from Quantum Kinetic Approach}, \href{https://journals.aps.org/prl/abstract/10.1103/PhysRevLett.109.232301}{Phys. Rev. Lett. {\bf 109}, 232301 (2012)}, [arXiv:1203.0725].

\bibitem{Gynther2011} Gynther Antti, Karl Landsteiner, Francisco Pena-Benitez, and Anton Rebhan. \textit{Holographic Anomalous Conductivities and the Chiral Magnetic Effect}, \href{https://doi.org/10.1007/JHEP02(2011)110}{J. High Energy Phys. {\bf 2011}, 110 (2011)}, [arXiv:1005.2587]. 

\bibitem{Bu2016a} Yanyan Bu, Michael Lublinsky, and Amir Sharon. \textit{Anomalous Transport from Holography. Part I} , \href{https://doi.org/10.1007/JHEP11(2016)093}{J. High Energy Phys. {\bf 2016}, 93 (2016) 11}, [arXiv:1608.0859]. 

\bibitem{Yee2009} Ho-Ung Yee.\textit{Holographic Chiral Magnetic Conductivity}, \href{https://doi.org/10.1088/1126-6708/2009/11/085}{J.High Energy Phys. {\bf 2009} (11), 085}, [arXiv:0908.4189].

\bibitem{Rebhan} Anton Rebhan, Andreas Schmitt, Stefan A. Stricker, \textit{Anomalies and the Chiral Magnetic Effect in Sakai-Sugimoto Model}, \href{https://doi.org/10.1007/JHEP01(2010)026}{J. High Energy Phys. {\bf 2010}, 26 (2010)}, [arXiv:0909.4782].

\bibitem{YJiang} Yin Jiang, Shuzhe Shi, Yi Yin, Jinfeng Liao, \textit{Quantifying Chiral Magnetic Effect from Anomalous-Viscous Fluid Dynamics} \href{https://iopscience.iop.org/article/10.1088/1674-1137/42/1/011001}{Chinese Phys. C {\bf 42}, 011001 (2018)}, [arXiv:1611.04586].

\bibitem{Shi} Shuzhe Shi, Hui Zhang, Defu Hou, Jinfeng Liao, \textit{Signatures of chiral magnetic effect in the collisions of isobars}, \href{https://journals.aps.org/prl/abstract/10.1103/PhysRevLett.125.242301}{Phys. Rev. Lett., {\bf 125}, 242301 (2020)}, [arXiv:1910.14010].

\bibitem{Guo} Xingyu Guo, Jinfeng Liao, Enke Wang. \textit{Spin Hydrodynamic Generation in the Charged Subatomic Swirl}. \href{https://doi.org/10.1038/s41598-020-59129-6}{Sci. Rep., {\bf 10}, 2196 (2020)}, [arXiv:1904.04704]. 
  
\bibitem{Rubakov2010} Rubakov, V. A. \textit{On Chiral Magnetic Effect and Holography}, \href{https://arxiv.org/abs/1005.1888}{ [hep-ph] arXiv:1005.1888}.

\bibitem{Miklos} Miklos Horvath, Defu Hou, Jinfeng Liao, Hai-cang Ren, \textit{Chiral magnetic response to arbitrary axial imbalance}, \href{https://journals.aps.org/prd/abstract/10.1103/PhysRevD.101.076026}{Phys. Rev. D {\bf 101}, 076026 (2020)}, [arXiv:1911.00933].

\bibitem{Feng} Bo Feng, De-fu Hou, Hai-cang Ren, Shuai Yuan, \textit{The noncommutativity of the static and homogeneous limit of the axial chemical potential in chiral magnetic effect}, \href{https://journals.aps.org/prd/abstract/10.1103/PhysRevD.103.056004}{Phys. Rev. D {\bf 103}, 056004 (2021)} \href{https://arxiv.org/abs/2008.10791}, [arXiv:2008.10791].

\bibitem{Adler} Stephen L. Adler, \textit{Axial-vector vertex in spinor electrodynamics} \href{https://journals.aps.org/pr/abstract/10.1103/PhysRev.177.2426}{Phys. Rev. {\bf 177}, 2426 (1969)}.

\bibitem{Adler_Bardeen} Stephen L. Adler and William A. Bardeen, \textit{Absence of higher-order corrections in the anomalous axial-vector divergence equation}, \href{https://journals.aps.org/pr/abstract/10.1103/PhysRev.182.1517}{Phys. Rev. {\bf 182}, 1517 (1969)}.

\bibitem{Coleman_Hill} Sidney Coleman, Brian Hill, \textit{No more corrections to the topological mass term in QED3}, \href{https://doi.org/10.1016/0370-2693(85)90883-4}{Phys. Lett. B {\bf 159}, 2-3 (1985)}.

\bibitem{Maldacena} Juan M. Maldacena, \textit{The Large $N$ Limit of Superconformal Field Theories and Supergravity}, \href{https://link.springer.com/article/10.1023/A:1026654312961}{Adv. Theor. Math. Phys. {\bf 38}, 1113-1133 (1998)}, [arXiv:hep-th/9711200]. 

\bibitem{Witten} Edward Witten, \textit{Anti-de Sitter space and Holography}, \href{https://dx.doi.org/10.4310/ATMP.1998.v2.n2.a2}{Adv. Theor. Math. Phys. {\bf 2}, 253 (1998)}, [arXiv:hep-th/9802150].  

\bibitem{Freedman} Daniel Z. Freedman, Samir D. Mathur, Alec Matusis, Leonardo Rastelli, \textit{Correlation functions in the $CFT_d/AdS_{d+1}$ correspondence}, \href{https://www.sciencedirect.com/science/article/abs/pii/S055032139900053X}{Nucl. Phys. B {\bf 546}, 96 (1999)}, [arXiv:hep-th/9804058].

\bibitem{Whittaker} E. T. Whittaker and G. N. Watson, \textit{A Course of Modern Analysis}, 4th ed. Cambrige University Press, 1927, Chapter XXIII.

\bibitem{Policastro2002a} Giuseppe Policastro, Dam T. Son and Andrei O. Starinets. \textit{From AdS/CFT Correspondence to Hydrodynamics}, \href{https://doi.org/10.1088/1126-6708/2002/09/043}{J. High Energy Phys. {\bf 2002}, 09 (2002)}, [arXiv:hep-th/0205052].

\bibitem{Bu2016} Yanyan Bu, Michael Lublinsky, Amir Sharon. \textit{U(1) Current from the AdS/CFT: Diffusion, Conductivity and Causality}, \href{https://doi.org/10.1007/JHEP04(2016)136}{J. High Energy Phys. {\bf 2016}, 136 (2016)}, [arXiv:1511.08789].

\bibitem{Bu2019} Yanyan Bu, Tuna Demircik, Michael Lublinsky. \textit{Nonlinear Chiral Transport from Holography}, \href{https://link.springer.com/article/10.1007/JHEP01(2019)078}{J. High Energy Phys. {\bf 2019}, 78 (2019)}, [arXiv:1807.08467].

\bibitem{Ansel'm}A. A. Ansel'm and A. A. Iogansen, \textit{Radiative corrections to the axial anomaly}, Zh. Eksp. Teor. Fiz. {\bf 96}, 1181 (1989) (English translation: \href{http://www.jetpletters.ac.ru/ps/1115/article_16872.shtml}{Sov. Phys. JETP {\bf 96}, 670 (1989)} ).

\end{thebibliography}
% 

\end{document}